\documentclass[aps,prl,twocolumn,showpacs,superscriptaddress,groupedaddress,nofootinbib,floatfix]{revtex4-1}

\pdfoutput=1 
\usepackage{amsmath,amstext}
\usepackage[T1]{fontenc}
\usepackage[utf8]{inputenc}
\usepackage{graphicx}
\usepackage[colorlinks=true,linktocpage=false,citecolor=blue,linkcolor=blue,urlcolor=blue]{hyperref}
\usepackage{multirow}
\usepackage[normalem]{ulem}
\usepackage{xr}
\usepackage{longtable}
\newcommand{\be}{\begin{equation}}
\newcommand{\ee}{\end{equation}}

\def\Msun{{\rm M_{\odot}}}

\def\Msun{{\rm M_{\odot}}}
\def\GMc2{{\rm G M_{\odot} c^{-2}}}

\def\abhi{a_\text{BH}}
\def\mbhi{M_\text{BH}}
\def\mns{M_\text{NS}}
\def\Rns{R_\text{NS}}
\def\abhf{a_{\bullet}}
\def\Mbhf{M_{\bullet}}
\def\mbhf{X_{\bullet}}

\usepackage{color}
\definecolor{cyan}{rgb}{0,0.9,0.9}
\definecolor{orange}{rgb}{0.9,0.5,0}
\definecolor{magenta}{rgb}{1,0,1}
\definecolor{purple}{rgb}{0.8,0.4,0.8}
\definecolor{gray}{rgb}{0.5,0.5,0.5}

\newcommand{\timesto}[1]{\times 10^{#1}}

\begin{document}

\title{Black-hole remnants from black-hole--neutron-star mergers}

\author{Francesco \surname{Zappa}${}^{1}$}
\author{Sebastiano \surname{Bernuzzi}${}^{1}$}
\author{Francesco \surname{Pannarale}${}^{2,3}$}
\author{Michela \surname{Mapelli}${}^{4,5,6,7}$}
\author{Nicola \surname{Giacobbo}${}^{4,5,6}$}
\affiliation{${}^1$Theoretisch-Physikalisches Institut, Friedrich-Schiller-Universit{\"a}t Jena, 07743, Jena, Germany}  
\affiliation{${}^2$Dipartimento di Fisica, Universit\`a di Roma “Sapienza”, Piazzale A. Moro 5, I-00185, Roma, Italy}  
\affiliation{${}^3$ INFN Sezione di Roma, Piazzale A. Moro 5, I-00185, Roma, Italy}
\affiliation{${}^4$Physics and Astronomy Department Galileo Galilei, University of Padova, Vicolo dell’Osservatorio 3, I–35122, Padova, Italy}
\affiliation{${}^5$INAF-Osservatorio Astronomico di Padova, Vicolo dell’Osservatorio 5, I–35122, Padova, Italy}
\affiliation{${}^6$INFN-Padova, Via Marzolo 8, I–35131 Padova, Italy}
\affiliation{${}^7$Institut f\"ur Astro- und Teilchenphysik, Universit\"at Innsbruck, Technikerstrasse 25/8, A–6020, Innsbruck, Austria}

\begin{abstract}
  Observations of gravitational waves and their electromagnetic
  counterparts may soon uncover the existence of coalescing compact
  binary systems formed by a stellar-mass black hole and a neutron
  star.  These mergers result in a remnant black hole,
  possibly surrounded by an accretion disk.  The mass and spin of the
  remnant black hole depend on the properties of the coalescing
  binary.
  We construct a map from the binary components to the remnant black
  hole using a sample of numerical-relativity simulations of
  different mass ratios $q$, (anti-)aligned dimensionless
  spins of the black hole $\abhi$, and several neutron star
  equations of state.
  Given the binary total mass, the mass and spin of the remnant black
  hole can therefore be determined from the three parameters
  $(q,\abhi,\Lambda)$, where $\Lambda$ is the tidal deformability of
  the neutron star. Our models also incorporate the binary black
  hole and test-mass limit cases and we discuss a simple extension for
  generic black hole spins.
  We combine the remnant characterization with recent population
  synthesis simulations for various metallicities of the progenitor
  stars that generated the binary system. We predict that black-hole--neutron-star mergers produce a population of remnant black
  holes with masses distributed around $7M_\odot$ and $9M_\odot$.
  For isotropic spin distributions, nonmassive accretion disks are favoured: 
  no bright electromagnetic counterparts are expected in such mergers.
\end{abstract}

\maketitle

\paragraph*{Introduction.---}
Mergers of a stellar-mass black hole (BH) and a neutron star (NS),
hereafter BHNS, are expected sources of gravitational waves (GWs)
detectable by ground-based laser-interferometers and possibly
accompanied by electromagnetic counterparts~\cite{Rosswog:2005su,
  Kyutoku:2011vz, Foucart:2012vn, Foucart:2014nda, Foucart:2015gaa,
  Paschalidis:2016agf, Bhattacharya:2018lmw}. No GW observations of 
BHNS binaries have been made to date. The 90\% confidence upper limit
on their merger rate is $610~{\rm Gpc}^{-3}{\rm
  yr}^{-1}$~\cite{TheLIGOScientific:2018mvr}. 
To prepare these observations quantitative general-relativistic
theoretical models of the GW and merger outcome are required.

Numerical-relativity (NR) simulations of BHNSs are the only means to 
study BHNS mergers
~\cite{Kyutoku:2010zd, Kyutoku:2011vz, Kyutoku:2013wxa,
  Kyutoku:2015gda, Foucart:2010eq, Foucart:2012vn, Foucart:2014nda,
  Kawaguchi:2015bwa, Etienne:2008re, Etienne:2007jg, Etienne:2011ea,
  Etienne:2012te, Paschalidis:2013jsa}. Simulations 
indicated that the NS tidal disruption is a characteristic
feature of the dynamics of quasi-circular BHNS mergers. On the contrary,
quasi-circular binary NS mergers with mass ratio $q\lesssim2$ do not present
significant tidal disruption, e.g.,~\cite{Bejger:2004zx, Dietrich:2016hky}.
Physically, tidal disruption
is expected if the binary reaches a characteristic radius
$r_\text{TD}$ before the innermost stable circular orbit
(ISCO), the radius of which we denote by
$r_\text{ISCO}$.  $r_\text{TD}$ is expected to scale in the same way
as the mass-shedding radius $r_\text{MS}$, which is determined by the
condition that the BH tidal force overcomes the NS self-gravity
at the stellar surface $\Rns$: $r_\text{TD}\lesssim r_\text{MS}\propto
q^{1/3} \Rns$, with a weak dependency on the BH
spin~\cite{Shibata:2011jka}.  For a Kerr BH of mass $\mbhi$, 
$r_\text{ISCO}=\mbhi f(\abhi)$, where
$f(\abhi)\in[1,9]$ is a monotonically decreasing function of the BH 
dimensionless spin parameter $\abhi$~\footnote{The dimensionless
  parameter range is $\abhi\in[-1,1]$ accounting for anti- and aligned spins.}
~\cite{Bardeen:1972fi}.
 Because $\Rns/\mbhi=(qC)^{-1}$, where $C=M_\text{NS}/\Rns$ is the NS compactness, the ratio that regulates the onset of
tidal disruption is 
$\xi=r_\text{TD}/r_\text{ISCO}\propto C^{-1} q^{-2/3} f(\abhi)^{-1}$.  Thus, tidal
disruption depends on three physical parameters: the binary mass
ratio, the BH spin and the NS compactness.

Simulations have
shown that tidal disruption occurs for BHNSs with $q \lesssim 3$ if the BH is
non-spinning, or its spin is anti-aligned with the orbital angular
momentum. Generally speaking, large, aligned BH spins $\abhi\gtrsim
+0.5$ favour tidal disruption because spin-orbit interactions push the
ISCO radius to smaller values.  As an example,
$r_\text{ISCO}=1\,M_\text{BH}$ for a Kerr BH with $\abhi=+1$, as
opposed to $r_\text{ISCO}=6\,M_\text{BH}$ for a non-spinning BH.
Disruption is also favored by low values of the NS compactness, which
are related to stiff equations of state, that also imply large NS
tidal deformabilities~\cite{Hinderer:2009ca, Damour:2009wj}. Note
that, for a fixed NS mass, large deformabilities imply large NS radii and
small binary mass ratios correspond to small BH masses.

\begin{figure*}[t]
  \centering 
  \includegraphics[width=0.95\textwidth]{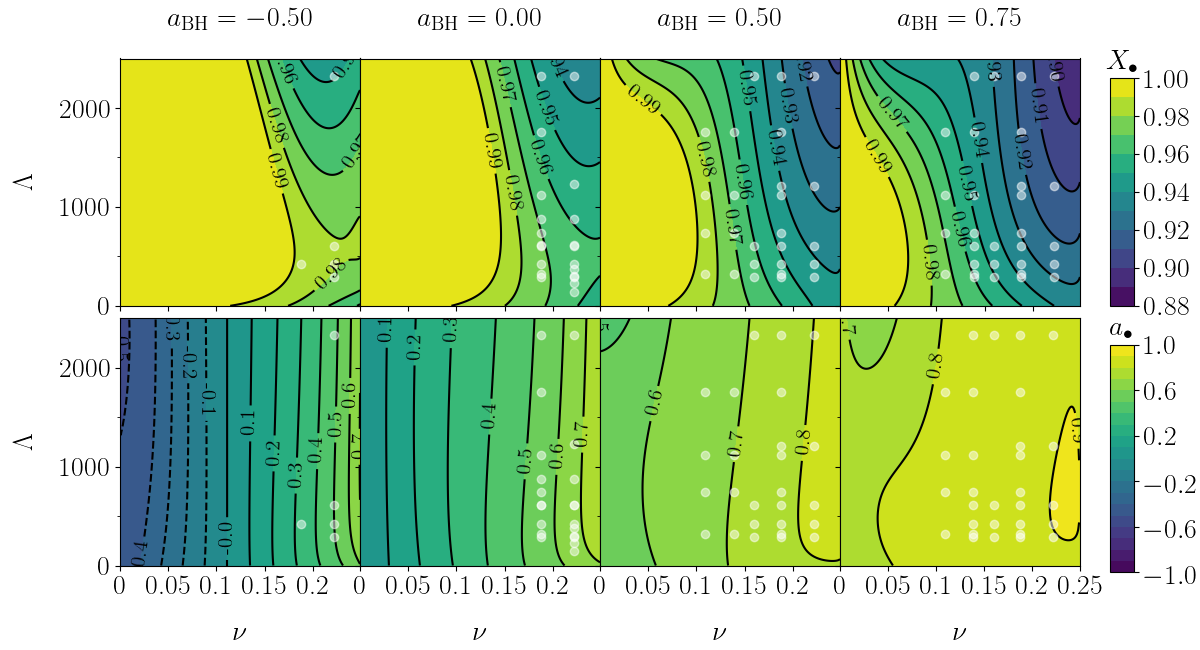}
  \caption{%
    Contour plots of the remnant BH mass divided by the binary mass
    $\mbhf=\Mbhf/M$ (top) and of the dimensionless spin parameter
    $\abhf$ (bottom) as a function of the symmetric mass ratio $\nu$
    and of the NS tidal polarizability parameter $\Lambda$, at fixed
    values of the initial BH spin parameter $\abhi$. The values of
    $\abhi$ correspond to those of the NR simulations.  White markers
    indicate the NR data used to construct the model.}
  \label{fig:fitmain}
\end{figure*}

Tidal disruption leads to the formation of an accretion disk in the
merger remnant. Simulations predict remnant disks with
baryon (rest) masses 
as large as $\gtrsim0.1\Msun$~\cite{Foucart:2012nc,Foucart:2018rjc}, thus creating the conditions
to ignite a short gamma-ray burst (SGRB)~\cite{Paczynski:1986px,
  Eichler:1989ve, Li:1998bw}.  \citet{Kyutoku:2011vz} tentatively
classify the phenomenology of BHNS mergers into three classes,
based on the ratio $\xi$.
For Type-I and Type-III mergers,
tidal disruption occurs far from or close to the ISCO; for Type-II it
does not occur and the NS plunges onto the BH, because the tidal
disruption radius is located well within the ISCO.  The three
classes differ by their GW spectra and the disk masses.
Type-II mergers are typically
characterized by $q\gtrsim 3,~\abhi\lesssim 0$ and have a GW spectrum
very similar to binary black holes (BBHs), e.g.~\cite{Lackey:2011vz,
  Pannarale:2013uoa, Pannarale:2015jka, Hinderer:2016eia,
  Nagar:2018zoe}.

An analytical formula for the BH remnant mass and dimensionless spin 
can be found using mass and angular momentum
conservation arguments~\cite{Pannarale:2012ux, Pannarale:2013jua} (see
also~\cite{Buonanno:2007sv}).  That approach builds on estimates of
the radiated energy and the binary orbital angular momentum based on
the expressions for test particles on Kerr background at ISCO, and on
the disk mass fits of \cite{Foucart:2012nc}.  Results are accurate to
a few percent, which is comparable to the energy radiated in GWs. The
largest uncertainty comes from the disk mass estimates in 
simulations e.g. \cite{Foucart:2018rjc,Radice:2018pdn}.

In this work we model the remnant of BHNS using NR data. Using a
state-of-art synthetic population we predict that the most
likely BHNS mergers are of Type-II, leading to a population of light remnant
BHs. Throughout this work we use geometric units $c=G=1$ unless otherwise stated.

\paragraph*{Remnant mass and spin.---}
Given the gravitational binary mass $M=\mbhi+\mns$, we map the remnant mass and spin
parameters of BHNS mergers as follows: 
\be\label{eq:F}
F: (\nu,\abhi,\Lambda) \to (\mbhf,\abhf)\ .
\ee
where $\mbhf=\Mbhf/M$ and $\abhf=S_\bullet/\Mbhf^2$, $\Mbhf$ and
$S_\bullet$ being the mass and spin of the remnant BH, respectively.
Above, $\nu=q/(1+q)^2\in[0,1/4]$ is the symmetric mass ratio ($q=
\mbhi/\mns\geq1$), spanning from the test-mass ($\nu=0$) to the
equal-mass ($\nu=1/4$) limit. $\abhi$ is the
dimensionless spin of the initial BH, that is aligned with the binary
orbital angular momentum. The quantity $\Lambda$ is the NS
quadrupolar tidal polarizability dimensionless
parameter~\cite{Hinderer:2009ca},
$\Lambda = 2k_2/(3C^5)$, 
where $k_2$ is the gravito-electric quadrupolar Love number, a
monotonically decreasing function of the compactness $C$~\cite{Damour:2009wj}.
$\Lambda$ describes tidal interactions at the leading order in
post-Newtonian dynamics.  Typically, $\Lambda\sim 100-2500$
for NS in BHNS systems, depending on the NS mass and equation of
state (EOS).

We use data of NR simulations of quasi-circular
BHNS mergers described in~\cite{Kyutoku:2010zd, Kyutoku:2011vz,
  Kyutoku:2015gda} and collected in the Supplementary Material (SM). These simulations adopt
different neutron star matter EOSs and (anti-)aligned BH 
spin values.

The NS spin on the contrary is neglected and currently not accounted for in our
models; however, this is expected to be a good approximation of realistic
systems~\cite{Bildsten:1992a, Kochanek:1992wk}.

The data cover the following parameter intervals: $\Lambda \in [100,
2500]$, $\nu \in [0.109, 0.222]$, and $\abhi \in [-0.5, 0.75]$. The
mapping $F$ is summarized in~\autoref{fig:fitmain}; technical details
on its construction are provided in the SM. 

The remnant BH mass scaled to $M$ is given by
\begin{equation}\label{eq:X}
  \mbhf = 1 - \frac{E_\text{GW}}{M} - \frac{M_\text{disk}}{M} \ ,
\end{equation}
where $E_\text{GW}$ is the
total energy radiated in GWs during the coalescence and
$M_\text{disk}$ is the disk contribution to the gravitational energy
which cannot be directly measured in the simulations~\footnote{In
  this formula for mass conservation, we neglect a term $M_\text{ejecta}\ll M_\text{disk}$ for the
  ejecta mass.}. In BBH 
mergers finite mass-ratio effects are repulsive, implying that the GW
emission is more efficient for larger $\nu$. The same effect is
present in the BHNS dynamics: \autoref{fig:fitmain} shows that the
smallest values of $\mbhf$ are obtained for larger values $\nu\to1/4$.
The precise behaviour of $\mbhf$, however, depends on the competition
between the energy emitted in GWs and the effect of tidal disruption,
as per Eq.~\eqref{eq:X}. For non-spinning BHNS binaries (second column
in \autoref{fig:fitmain}), one observes that the value of $\mbhf$
slightly increases with respect to the BBH case as $\Lambda>0^+$ and
for a given $\nu$.  Tidal disruption does not occur for
small values of $\Lambda\sim0$, so this effect is solely due to the fact
that tidal interactions are
attractive and reduce the emission of GWs with respect to the
$\Lambda=0$ case (i.e., $E_\text{GW}$ decreases so $\mbhf$ grows, with $M_\text{disk}\simeq 0$). As $\Lambda$ becomes sufficiently
large (and $\nu\to1/4$), tidal disruption occurs and only part of the
remnant mass contributes to the final BH mass. Consequently, as 
$\Lambda$ increases beyond a certain critical value,
$\mbhf$ starts to decrease
because part of the NS mass is not
swallowed by the BH but becomes part of the disk. 
Note that the peak mass is more pronounced for $\nu\to1/4$ and disappears for 
sufficiently small $\nu$ (Type-II mergers).

Focusing on spin effects, at a given $\nu$, the remnant mass decreases
for increasing $\abhi>0$ because the ratio $\xi$ 
increases. This is a consequence of the repulsive character of the
spin-orbit interaction for aligned (positive) spins.  Notably, the peak
for small $\Lambda$ is no longer present for sufficiently large values
of $\abhi$.  For $\abhi<0$, the spin-orbit interactions are
attractive, i.e., they have the same sign as tidal interactions. As a
consequence, for smaller $\abhi$'s, $\mbhf$ increases and the peak at
small $\Lambda$ is more pronounced.

For non-spinning BBHs, the remnant BH spin $S_\bullet$ 
is expected to decrease 
for increasing $\nu$, due to the same finite mass-ratio effect described
above. Due to the $M^2_\bullet$ normalization, however, $\abhf$ shows the
opposite behaviour.  In the BHNS case, the remnant BH has a larger
dimensionless mass-rescaled spin with respect to the BBH case and it
increases with $\Lambda$, for small $\Lambda>0$. This happens
because the NS compactness is smaller and less angular momentum is
dissipated via GWs. Above a peak value, however, tidal disruption
occurs and the angular momentum redistributes into the disk that forms
around the remnant BH.

For $|\abhi|\lesssim0.5$ and a given value of $\nu$, the final $\abhf$
is roughly linear in $\abhi$ [see Eq.~\eqref{eq:Ffit} and
\cite{Kyutoku:2011vz}].
For $\abhi\gtrsim+0.75$, one recovers $\abhf\sim a_\text{BBH}$, as expected.

Although our models are developed from non-precessing BHNS data,
they 
can be extended to the case of generic BH spins~\cite{Barausse:2009uz,
  Pannarale:2013jua, Hofmann:2016yih}.  The simplest extension ---
which we adopt --- is to map the initial spin
\begin{equation}
  \abhi \rightarrow \abhi \cos\beta = \abhi^z\ ,
  \label{eq:precessing}
\end{equation}
where $\beta$ is the angle between the initial BH spin and the orbital
angular momentum $\textbf{L}$. In this case the model will yield $\abhf^z$
instead of $\abhf$.
This prescription also assumes that the
direction of the total angular momentum $\textbf{J} = \textbf{L} +
\textbf{S}$ is approximately preserved and so the direction $\theta$
of the final spin is given by the projection
$\cos\theta = \hat{\textbf{J}} \cdot \hat{\textbf{L}}$\,.
Predictions in the precessing case agree with the simulations of
\cite{Kawaguchi:2015bwa}, (see SM).

\begin{figure}[t]
  \centering
  \includegraphics[width=0.5\textwidth]{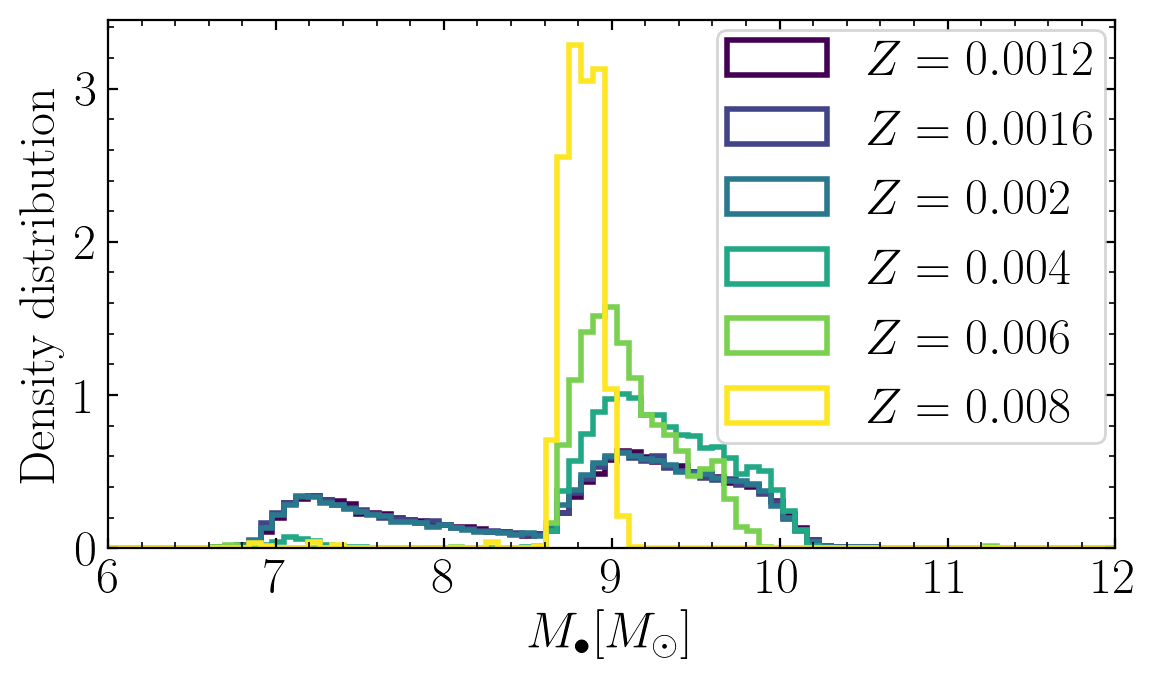}
  \caption{The remnant BH mass distribution inferred from the remnant to a different value 
  		   of the metallicity $Z$ of the progenitor stars. In this plot we employ the SLy EOS 
  		   and the fiducial isotropic spin distribution peaked around $\langle\abhi\rangle=0.2$.}
  \label{fig:mass_pop}
\end{figure}

\begin{figure*}[t]
  \centering
  \includegraphics[width=\textwidth]{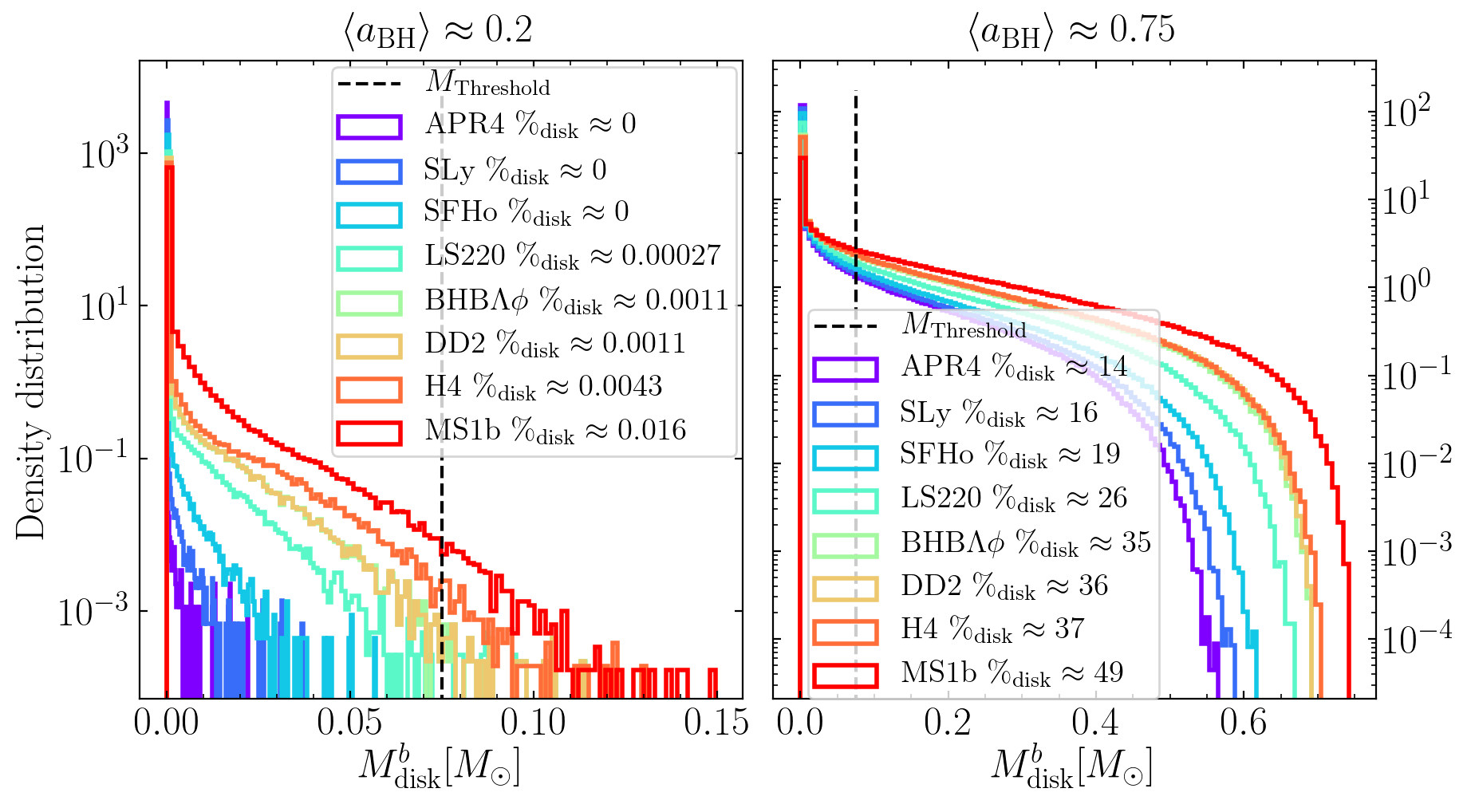}
  \caption{Remnant disk baryonic mass distribution for different EOSs
  		   and for low (left) and high (right) aligned BH spin distributions. The 
  		   mass threshold represents the minimum mass of the disk that allows the production 
  		   of SGRBs with $1$~s duration.
  		   The percentage of binaries with disk mass bigger than the 
  		   threshold is provided in the legend for each equation of state.}
  \label{fig:disk_summary}
\end{figure*}

\paragraph*{Binary and remnant population.---}
We now apply the formalism described in the previous section to a 
BHNS population merging at redshift $z\leq{}1$ and constructed by convolving the
binary population-synthesis from the {\sc{mobse}} code 
\cite{giacobbo2018a,giacobbo2018b,giacobbo2019} with the
Illustris cosmological simulation
(\cite{vogelsberger2014a,vogelsberger2014b,nelson2015}, see
\cite{mapelli2017,mapelli2018,mapelli2019} and the SM for details). 
In particular, we adopt run~CC15$\alpha{}5$ of \cite{mapelli2019}
where the common-envelope parameter is $\alpha{}=5$ and 
natal kicks are drawn from a Maxwellian distribution with a single root-mean square
velocity $v_{\sigma}=15$ km s$^{-1}$ for both electron-capture and
core-collapse supernovae. Larger kicks would enable the merger of more
massive BHNSs (moderate kicks do not break the binary but increase its eccentricity, shortening the
merger time of massive BHNSs, see \cite{giacobbo2018b} for details),
but would not affect the minimum BHNS mass (See
Figure~5 of \cite{mapelli2019}).  In run CC15$\alpha{}5$, the minimum (maximum) mass of a
BH (NS) is set to 5 (2)
$M_\odot$. This assumption enforces the existence of a mass gap between
BHs and NSs, which is suggested by dynamical
mass measurements of compact objects in X-ray binaries
\cite{ozel2010,farr2011}. 
BH spins are added by randomly drawing spin magnitudes
$|\abhi|\in [0,1]$ from a truncated Maxwellian distribution with root 
mean square $\sigma{}$. 
In this paper, we consider spins isotropically oriented with respect to the 
binary orbital plane with $\langle\abhi\rangle=0.2$ as fiducial distribution
or aligned spin distributions with $\sigma{}=(0.1, 0.35, 0.5, 0.7)$, corresponding to
average values $\langle\abhi\rangle=(0.2,0.5,0.75,0.95)$.
The aligned spin
distributions give {\it upper limits} to the isotropic spin distributions.

The population synthesis predicts BH component masses bellow
$10\Msun$ distributed narrowly about $\mbhi\sim 5\Msun$ and $\mbhi\sim
8\Msun$~\cite{Mapelli:2018wys}. The population depends very weakly on
progenitors' metallicities for $Z\leq0.002$, but for $Z\geq0.003$ the
smallest BHs are suppressed and only BH with $\mbhi\sim
8\Msun$ are found. 
This is a consequence of the dependence of 
the delay time (i.e., the time elapsed between the formation of the progenitor stars 
and the BHNS merger) on the progenitor's metallicity: metal-rich progenitors have longer
delay times than metal-poor ones and thus do not merge within the Hubble time, especially
if the BH mass is small [Giacobbo et al., In prep.]. 
Additionally, NS masses $\mns\gtrsim 1.3\Msun$ are favoured.

In order to compute the merger remnant from the population, we choose a
representative set of EOSs, and calculate $\Lambda$ on the NS population
for each EOS.
The remnant properties are then determined with Eq.~\eqref{eq:F} with 
the prescription of Eq.~\eqref{eq:precessing}. Remnant masses
are shown in~\autoref{fig:mass_pop}, while additional plots are reported in the SM. 
For metallicities $Z\leq0.002$, we find a bimodal distribution around
$\Mbhf\sim7\Msun$ and $\Mbhf\sim9\Msun$ independently from the
EOS. Large metallicities produce only the more massive remnants.
The remnant spins inferred from Eq.~\eqref{eq:F} and the
isotropic/aligned spin population with $\langle\abhi\rangle \approx 0.2$ 
are distributed 
around $\abhf^z\sim0.4$ with standard deviation $\sim 0.1$. 

Using the model of \cite{Foucart:2018rjc}, we estimate the baryonic 
mass of the remnant disk. \autoref{fig:disk_summary} shows the aligned 
low spin distribution resulting in $\gtrsim 99\%$ 
of the remnants with baryonic mass of the disk smaller than
$M^b_\text{Threshold}=0.075(\mns^b/1.5)\Msun$ independently from the EOS.
Disk masses above $M^b_\text{Threshold}$ are necessary to produce 
SGRBs of $1$~s duration~\cite{Stone:2012tr,Pannarale:2014rea}.
Remnants with significant disk masses are found for aligned
spin distributions with $\langle\abhi\rangle\gtrsim0.5$.
In these cases, the largest disks are found for the stiff EOS 
corresponding to $\Lambda\gtrsim 1700$.
Soft EOS, corresponding to $\Lambda\lesssim 400$,
give massive disks only for $\lesssim20\%$ of the binaries and with
$\langle\abhi\rangle\gtrsim0.75$. 

\paragraph*{Conclusion.---}
Our results indicate light and
moderately spinning BH remnants surrounded by low-mass 
accretion disks (Type-II) as the most likely
outcome for BHNS if $\Lambda\lesssim 1000$ and the BH has aligned
spin $\abhi\lesssim0.75$.
The observation of GW170817 rules out NS
with $\Lambda\gtrsim1800$ ($\gtrsim 2600$) for the low- (high-) spin prior cases \cite{PhysRevX.9.011001}. 
Similarly, large aligned spins might be disfavoured by current GW binary observations 
\cite{TheLIGOScientific:2018mvr}.
Type-II GW signals are very similar to BBHs. For aligned spins, 
GW searches will lose less than 1\% of
events employing BBH templates \cite{Harry:2016ijz}. 
On the other hand, estimating $\Lambda$ from the GW will be
challenging, and BHNS mergers might not set constraints on the
EOS unless ringdown signatures are resolved \cite{Pannarale:2012ux}.
Type-II mergers are also not expected to be accompanied by bright 
electromagnetic counterparts. Disk masses above 
$M^b_\text{Threshold}$ are rare in our populations, unless BHNSs are
characterized by large and aligned BH initial spins, very stiff EOS
and/or compact objects with mass $2-5\,{}M_\odot$ (i.e. within the mass gap suggested by X-ray binaries).     

The BH remnant model constructed in this work will be used in GW
models for BHNSs 
\cite{Lackey:2011vz,Lackey:2013axa,Pannarale:2013jua,Pannarale:2015jka,Pannarale:2015jia,Hinderer:2016eia,Nagar:2018zoe},  
as well as for modeling the counterparts,
e.g.~\cite{Lee:1999se,Beloborodov:2000nn,Kawaguchi:2016ana,Perego:2017wtu,Barbieri:2019sjc}.  
It will thus be one of the key building blocks for upcoming
multi-messenger analysis of BHNSs. \\

\acknowledgments
We thank Koutarou Kyutoku for discussions and for sharing the NR data used
in this work.
FZ and SB acknowledges support by the EU H2020 under ERC Starting
Grant, no.~BinGraSp-714626.  
FZ and FP acknowledge support from Cardiff University Seedcorn Funding AH21101018.
MM acknowledges financial support by the European 
Research Council for the ERC Consolidator grant DEMOBLACK, under contract no. 770017


\bibliography{references.bib,local.bib}

\onecolumngrid

\pagebreak
\widetext
\begin{center}
\textbf{\large Supplementary Materials}
\end{center}


\section{Data and model construction}
\label{app:fit}

We construct a model for the mass and spin of the remnant BH based on
the following expression
\be%
\label{eq:Ffit}%
F(\nu, \abhi, \Lambda)= 
F_\text{BBH}(\nu, \abhi)
\frac{1+p_1(\nu, \abhi)\Lambda+p_2(\nu,\abhi)\Lambda^2}
{\left(1+[p_3(\nu, \abhi)]^2\Lambda\right)^2}
 \ ,
\ee
where $F_\text{BBH}$ are the BBH models for mass and spin developed
in~\cite{Jimenez-Forteza:2016oae} and $p_k(\nu,\abhi)$ are polynomials
of the form
\begin{align}
p_{\rm k}(\nu, \abhi) &= p_{k1}(\abhi)~\nu + p_{k2}(\abhi)~\nu^2 \\
p_{kj}(\abhi) &= p_{kj0}~\abhi + p_{kj1} \ .
\end{align}
The model includes by construction the BBH limit for $\Lambda
\rightarrow0$ (no tidal effects) and the test-mass limit for $\nu
\rightarrow 0$.  Note that the dependence on BH spin is linear.
The coefficient $p_3(\nu, \abhi)$ is squared in order to avoid a pole
in the denominator.

\begin{figure}[t]
  \centering
  \includegraphics[width=.48\textwidth]{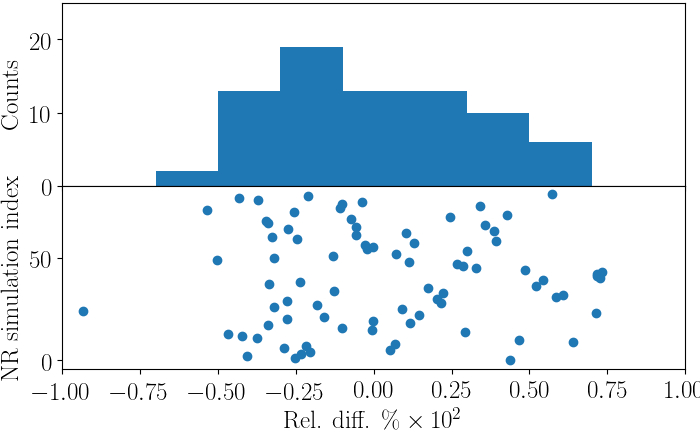}
  \includegraphics[width=.48\textwidth]{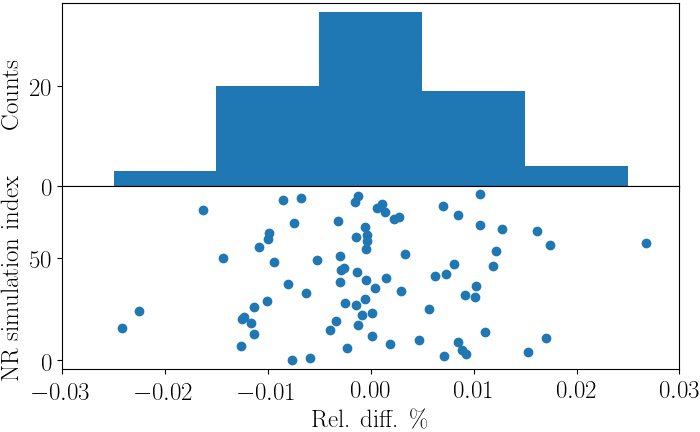}
  \caption{Relative differences of the data points to the fit. Left panel:
    $X_\bullet$, the maximum difference is of the order of
    $1\%$. Right panel: $a_\bullet$, the maximum difference is below
    $3\%$. In both cases the coefficient of determination of the fit
    is $R^2 \sim 0.92$.}
  \label{fig:res}
\end{figure}

The NR data used for our work is collected in~\autoref{tab:data}.
The values of the best-fit parameters $p_{kjl}$ are reported in
\autoref{tab:par_fit}; relative differences with respect to the 
fit are shown
in \autoref{fig:res}. The fits have determination coefficient
$R^2\sim 0.92$ and the residuals are normally distributed with
mean $\sim 0$ and standard deviations $\sim 0.25\times10^{-2}$ and $\sim 0.01$, respectively. 
The maximum relative
differences are below $1\%$ for the remnant mass fit and below $3\%$
for the the remnant spin; hence, fit uncertainties are smaller than NR
errors. The NR data do not extend to $\nu\lesssim0.1$ and 
$\abhi\lesssim-0.5 \vee \abhi\gtrsim0.75$, thus the
model effectively extrapolates into those regions. The extrapolation
leads in some cases to unphysical values of $\mbhf>1$ and $\abhf < -1$. 
This behaviour is fixed by forcing the model to agree
with the BBH, i.e., forcing $\mbhf=1$ and $|\abhf| \leq 1$.
We expect that future
calibration of the model with new NR data will improve the behaviour
for large mass ratios. We stress that BHNSs with mass ratios
$q\gtrsim7$ are in any case effectively indistinguishable from BBHs.

\begin{figure}[t]
  \centering
  \includegraphics[width=\textwidth]{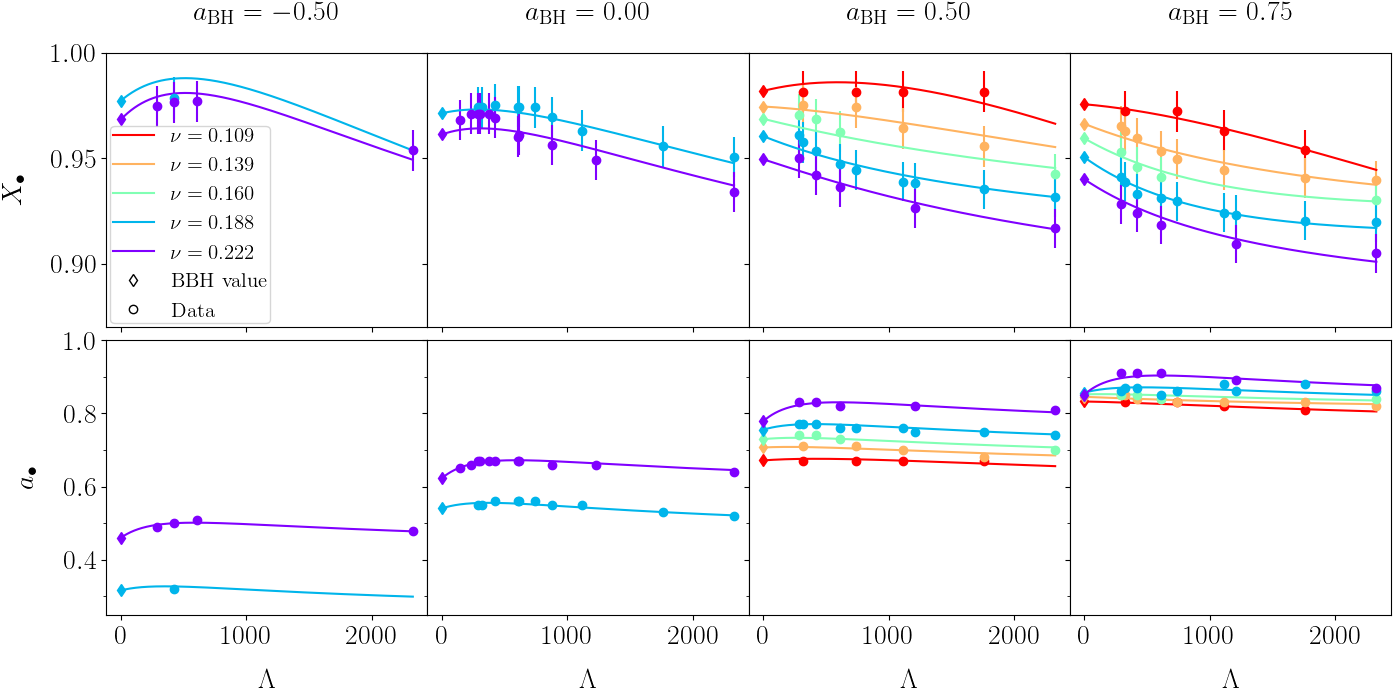}
  \caption{The remnant BH fractional mass $\mbhf=\Mbhf/M$ (top) and
    dimensionless spin parameter $\abhf$ (bottom) as
    a function of the tidal polarizability parameter $\Lambda$ at
    given values of the initial BH spin parameters $\abhi$.
    The values of $\abhi$ correspond to those of the NR simulations.}
  \label{fig:fit1d}
\end{figure}

\begin{figure}[t]
  \centering
  \includegraphics[width=\textwidth]{fig01.png}
  \includegraphics[width=\textwidth]{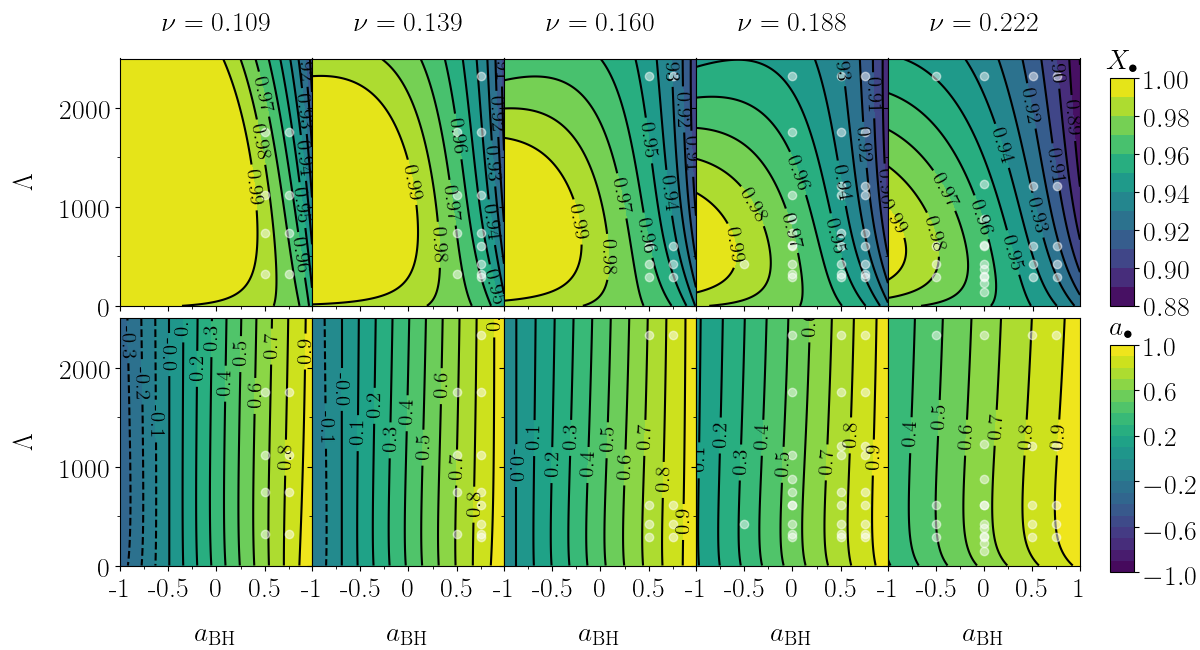}
  \caption{%
    Top: Contour plots of the remnant BH fractional mass $\mbhf=\Mbhf/M$
    and dimensionless spin parameter
    $\abhf$ as functions of the symmetric mass ratio $\nu$ and the NS
    tidal polarizability parameter $\Lambda$ at given values of
    the initial BH spin parameters $\abhi$.
    The values of $\abhi$ correspond to those of the NR simulations.
    Bottom: The same physical quantities are shown
    as functions of the initial BH spin parameters $\abhi$ and the NS
    tidal polarizability parameters $\Lambda$ at given values of
    the symmetric mass ratio $\nu$. The values of $\nu$
    correspond to those of the NR simulations. 
    White markers indicate the NR data used to construct the model.
  }
  \label{fig:fit2d}
\end{figure}

\autoref{fig:fit1d} and \autoref{fig:fit2d} complement
\autoref{fig:fitmain} in the main text. \autoref{fig:fit1d} shows
the model dependencies on $\Lambda$ for each given value of
$\abhi$. We show NR errors in the plot, optimistically taken to be at
the $1\%$ level. Note that in the final spin plot the errorbars 
are smaller than the markers.  The contour plots in
\autoref{fig:fit2d} show the model evaluated on the entire
parameter space (the plots in the main section of the paper are here
reproduced for completeness).

\begin{table}[t]
  \caption{Best fit parameters of the remnant model for mass, spin and
    peak luminosity and the respective determination coefficient $R^2$ of the fit.}         
    \label{tab:par_fit}
\begin{tabular}{ccccccc}
	\hline
	\hline    
    $F$ & $k$ & $p_{k10}$ & $p_{k11}$ & $p_{k20}$ & $p_{k21}$ & $R^2$ \\
    \hline
    & $1$ & $-1.83\timesto{-3}$ & $2.39\timesto{-3}$ & $4.29\timesto{-3}$ & $9.8\timesto{-3}$ & \\
    $\mbhf$ & $2$ & $2.34\timesto{-7}$ & $-8.28\timesto{-7}$ & $-1.64\timesto{-6}$ & $8.08\timesto{-6}$ & $0.921$\\
    & $3$ & $-2.01\timesto{-2}$ & $1.32\timesto{-1}$ & $6.51\timesto{-2}$ & $-1.43\timesto{-1}$ & \\
    \hline
    & $1$ & $-5.44\timesto{-3}$ & $7.91\timesto{-3}$ & $2.33\timesto{-2}$ & $2.48\timesto{-2}$ & \\
    $\abhf$ & $2$ & $-8.57\timesto{-7}$ & $-2.82\timesto{-6}$ & $6.61\timesto{-6}$ & $4.29\timesto{-5}$ & $0.92$\\
    & $3$ & $-3.04\timesto{-2}$ & $2.55\timesto{-1}$ & $1.48\timesto{-1}$ & $-4.28\timesto{-1}$ & \\
    \hline
    & $1$ & $3.08\timesto{-2}$ & $-4.18\timesto{-2}$ & $-5.17\timesto{-2}$ & $3.19\timesto{-1}$ & \\
    $L_\text{peak}$ & $2$ & $-1.23\timesto{-5}$ & $8.84\timesto{-6}$ & $1.05\timesto{-4}$ & $-3.88\timesto{-5}$ & $0.98$\\
    & $3$ & $3.30\timesto{-1}$ & $-3.76\timesto{-2}$ & $-9.2\timesto{-1}$ & $1.44$ & \\
    \hline
    \hline
\end{tabular}
\end{table}

\begin{figure*}[t]
  \centering
  \includegraphics[width=\textwidth]{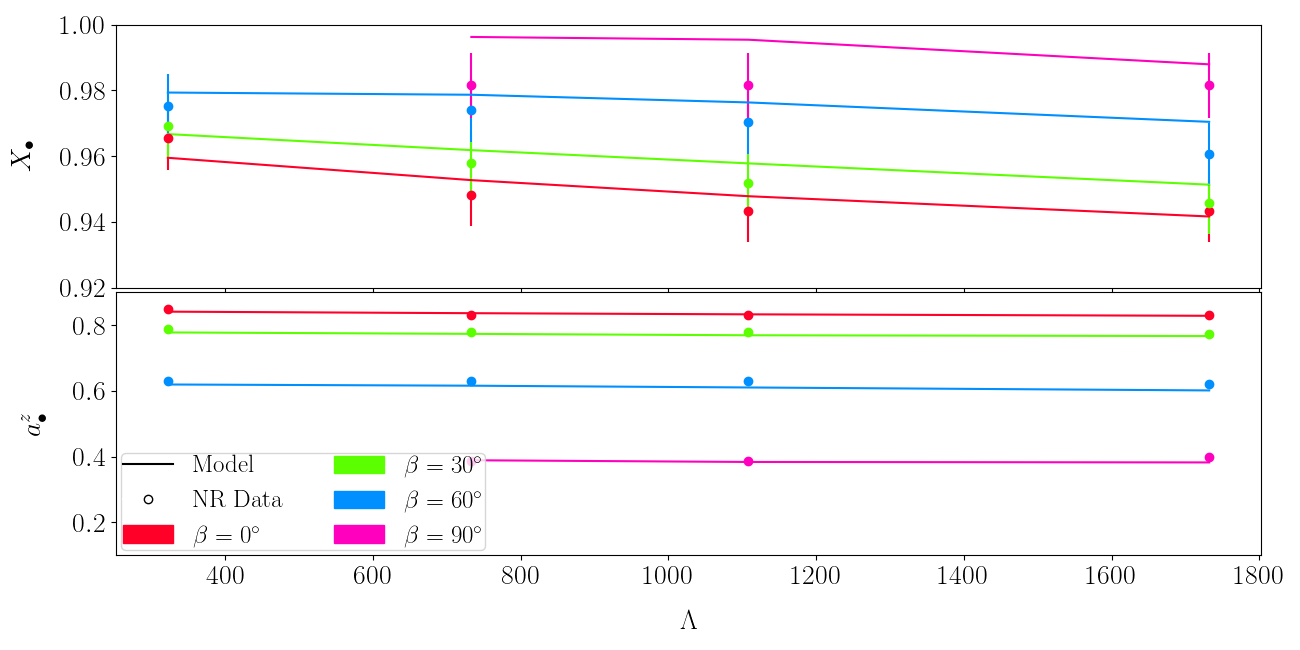}
  \caption{Model performance for binaries with
    spin-precession. The model is validated against the numerical data
    of \cite{Kawaguchi:2015bwa}. $\beta$ is the angle between the initial
    orbital angular momentum and the initial BH spin. Note that the exact values of $\beta$
	are slightly different from the ones reported here, as explained in the
	reference. All these data correspond to fixed values of $(\abhi,q) = (0.75, 5)$ and the
	errorbars are taken as a $1\%$ of the values.}  
  \label{fig:prec_res}
\end{figure*}

Finally, we validate the prescription used to extend the model to spin
precessing cases. \autoref{fig:prec_res} shows the agreement between
the model predictions for $\mbhf$ and $\abhf$ using
Eq.~\eqref{eq:precessing} and the NR data of \cite{Kawaguchi:2015bwa}
(not used to determine the model).

\section{GW peak luminosity}

\begin{figure}[t]
  \centering 
  \includegraphics[width=\textwidth]{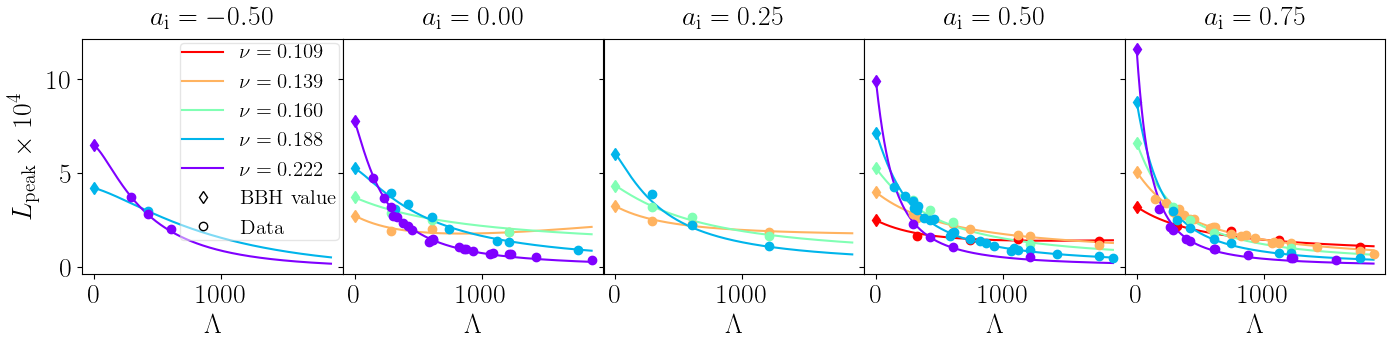}
  \caption{Distribution of the GW peak luminosity as a function of the
    NS $\Lambda$ for all the values of $\abhi$ and for all the
    different values of $\nu$.}
  \label{fig:lum_fit}
\end{figure}

The approach used in this paper can be used to estimate the GW peak
luminosity, thus complementing the result derived for BBHs and binary
NS systems in \cite{Keitel:2016krm, Zappa:2017xba} (see \cite{TheLIGOScientific:2018mvr} for an
application of those results).  The GW peak luminosity is computed
from the $(2,2)$ mode of the GW strain,
\begin{equation}
  L_\text{peak} = \max_t \frac{1}{16\pi} \left(\left|\frac{dh_{22}(t)}{dt}\right|^2\right)
\end{equation}
where 
\begin{equation}
  h_+ - i h_\times = \sum_{\ell, m} h_{\ell m}(t) {}^{-2}Y_{\ell, m}
  \approx h_{22}(t) ({}^{-2}Y_{2, 2} + {}^{-2}Y_{2, -2}) \ , 
\end{equation}
and ${}^{-2}Y_{\ell, m}$ are the spin-weighted $s=-2$ spherical harmonics.

\autoref{fig:lum_fit} shows the peak luminosity model as a function of
$\Lambda$ and for the values of $\nu$ and $\abhi$ sampled by the NR
dataset. The behaviour of $L_\text{peak}$ closely mirrors the one for
$\mbhf$ detailed in the main text.  The model for $L_\text{peak}$ is
slightly different from Eq.~\eqref{eq:Ffit}; in order to avoid
negative, unphysical values we fit the ansatz
\be%
\label{eq:Efit_lpeak}%
L_\text{peak}(\nu, \abhi, \Lambda) =%
L_\text{peak\, BBH}(\nu, \abhi)%
\frac{\left(1+p_1(\nu, \abhi)\Lambda+p_2(\nu,\abhi)\Lambda^2\right)^2}
{\left(1+[p_3(\nu, \abhi)]^2\Lambda\right)^4} \ .
\ee
The best parameters $p_{kjl}$ are reported in
\autoref{tab:par_fit}. The model delivers results accurate at the $20 \%$
level.

The python code which implements these models is available at \url{https://git.tpi.uni-jena.de/core/bhnsremnant}.

\section{Synthetic population}

The BHNS population is constructed from the results of the binary
population-synthesis code {\sc{mobse}}
\cite{giacobbo2018a,giacobbo2018b,giacobbo2019} convolved with the
Illustris cosmological simulation
\cite{vogelsberger2014a,vogelsberger2014b,nelson2015}, and following
the Monte Carlo method already described in \cite{mapelli2017,
  mapelli2018}. {\sc{mobse}} includes up-to-date prescriptions for
stellar winds (accounting for the stellar metallicity and luminosity
dependence of the mass loss), core-collapse supernovae (SNe),
electron-capture SNe and (pulsational) pair instability SNe (see
\cite{giacobbo2018b} for more details). The interface with the
Illustris simulation enables us to know the merger redshift of each
simulated BHNS (see \cite{mapelli2018}). We consider only BHNSs
merging at redshift $z\leq{}1$.

Here, we adopt run~CC15$\alpha{}5$ discussed in \cite{mapelli2019}. In
this run, we fix common-envelope parameter $\alpha{}=5$ and we draw
natal kicks from a Maxwellian distribution with one root-mean square
velocity $v_{\sigma}=15$ km s$^{-1}$ for both electron-capture and
core-collapse SNe. Larger kicks would enable the formation of few more
massive BHNSs, but would not affect the minimum BHNS mass (see
Figure~5 of \cite{mapelli2019}). The minimum (maximum) mass of a
BH (NS) in run CC15$\alpha{}5$ is set to 5 (2)
$M_\odot$ (see the rapid model in \cite{fryer2012} for
details). This assumption enforces the existence of a mass gap between
BHs and NSs, which is mildly suggested by dynamical
mass measurements of compact objects in X-ray binaries
\cite{ozel2010,farr2011}. 


BH spins are added in post-processing through a toy model as
there is no commonly accepted model to derive BH spins from 
the properties of their stellar progenitors. We randomly assign BH 
spin magnitudes $0\leq{}|\abhi|\le{}1$ from a truncated Maxwellian
distribution with one-dimensional root mean square $\sigma{}$.
For the analysis of the remnant black hole distribution we assume
spins isotropically oriented with respect to the binary
orbital plane with modulus $\langle\abhi\rangle \approx 0.2$ using
the prescription for precessing binaries described in the main text.
In the disk analysis, by contrast, we consider different aligned spin distributions 
with $\sigma{}=(0.1, 0.35, 0.5, 0.7)$ corresponding to means
$\langle\abhi\rangle=(0.2,0.5,0.75,0.95)$ because the forumlae we use
to estimate the disk mass are developed for aligned black hole spins. 
However, the aligned spin distributions give {\it upper limits} to the isotropic 
spin distributions. We assume all NSs have zero spins. 

\begin{figure*}[t]
  \centering
  \includegraphics[width=\textwidth]{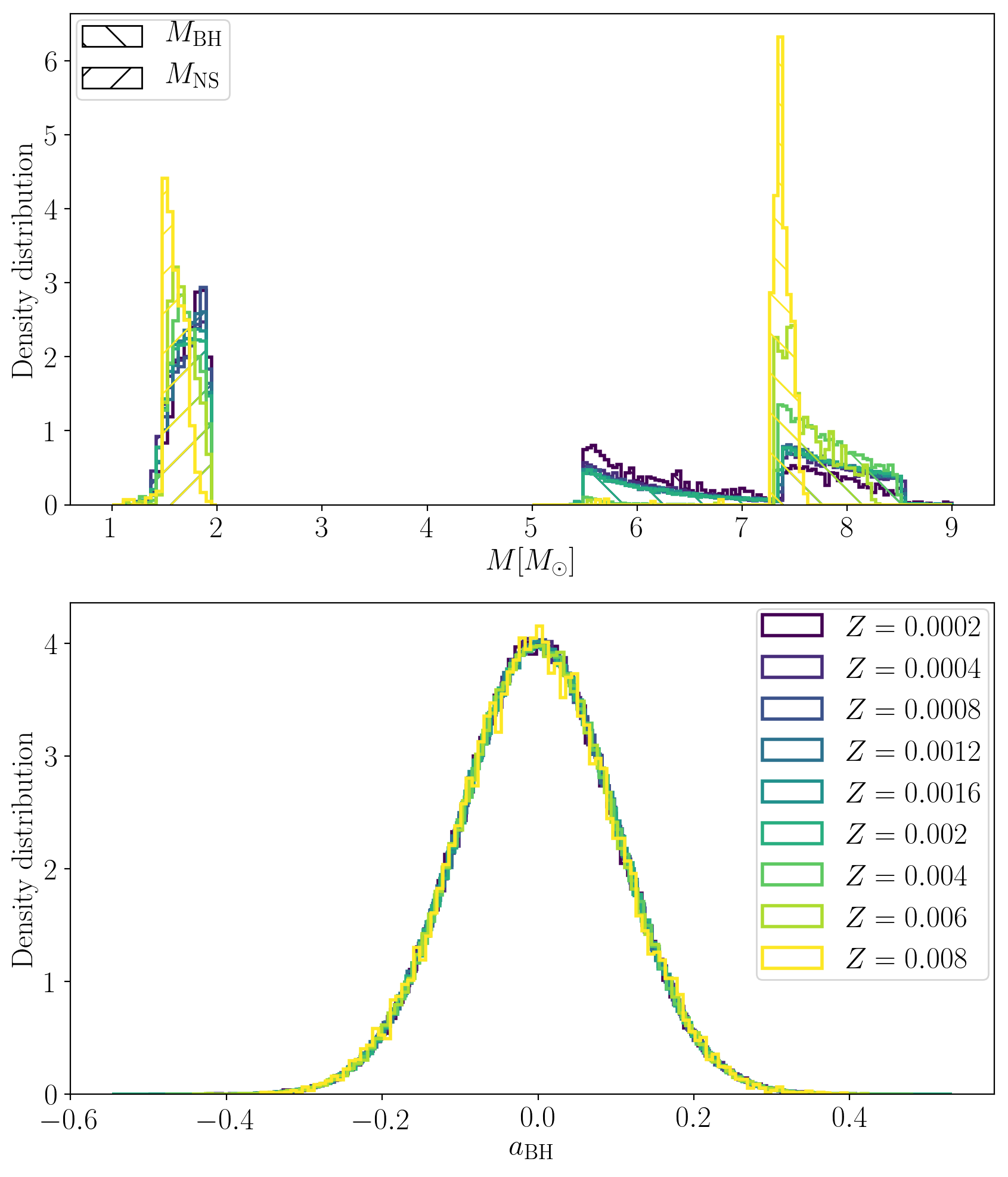}
  \caption{Distribution of initial component masses of the BHNS systems from population
  		   synthesis simulations (top panel) and initial BH spin (bottom panel)
  		   for different metallicities.}  
  \label{fig:pop_distributions}
\end{figure*}

\begin{figure*}[t]
  \centering
  \includegraphics[width=\textwidth]{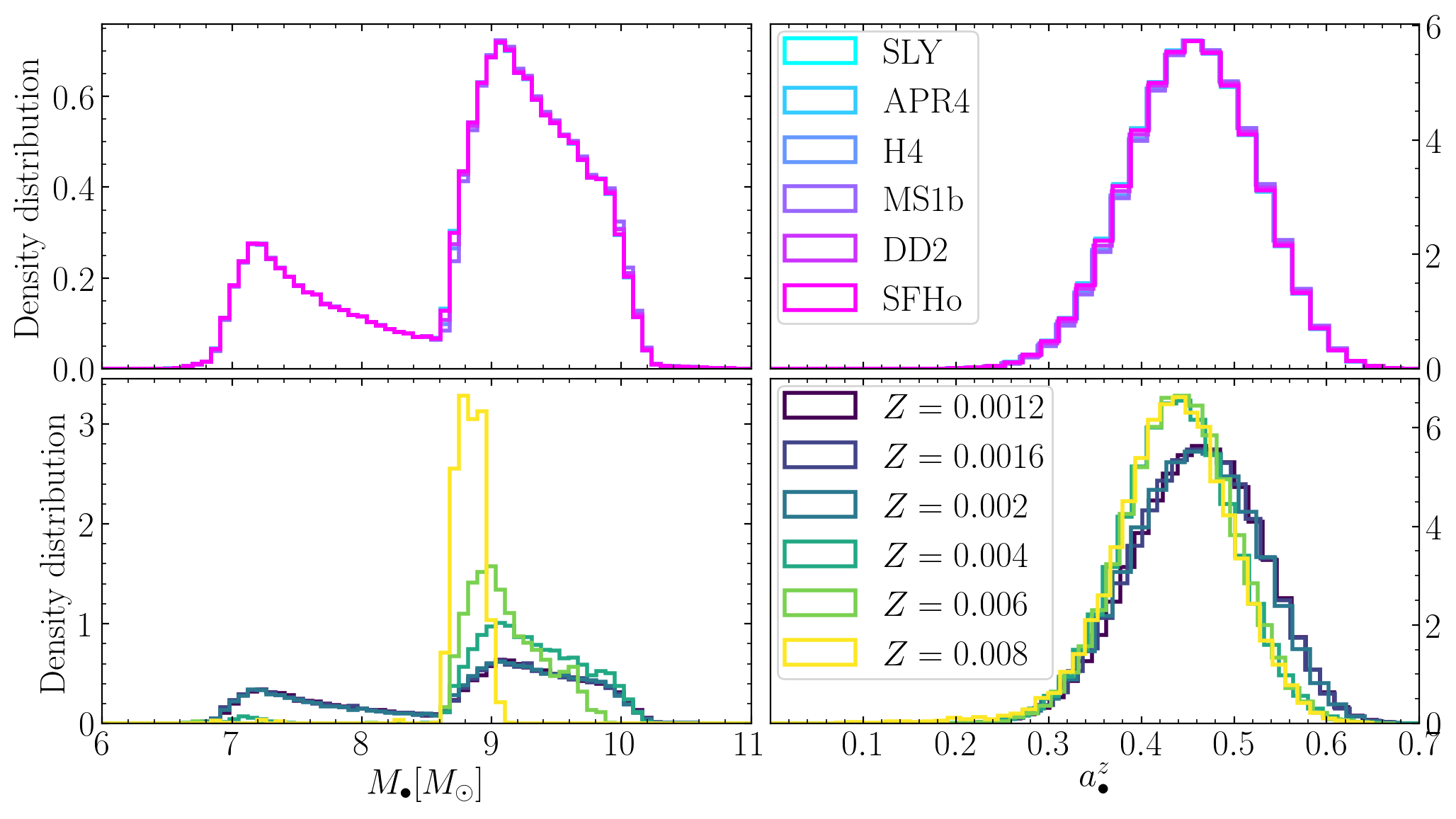}
  \caption{Final BH mass distribution (left) and final BH spin distribution (right)
  		   for different equations of state (top panels) and for different metallicities (bottom panels).
  		   Note that the prescription for precessing BHs is employed and so the $z-$component of
  		   the spin is reported.  In this plot we employ the fiducial isotropic spin distribution 
  		   peaked around $\langle\abhi\rangle=0.2$ and the SLy EOS is used in the bottom panels.}  
  \label{fig:pop_distributions_rem}
\end{figure*}

\begin{figure*}[t]
  \centering
  \includegraphics[width=0.49\textwidth]{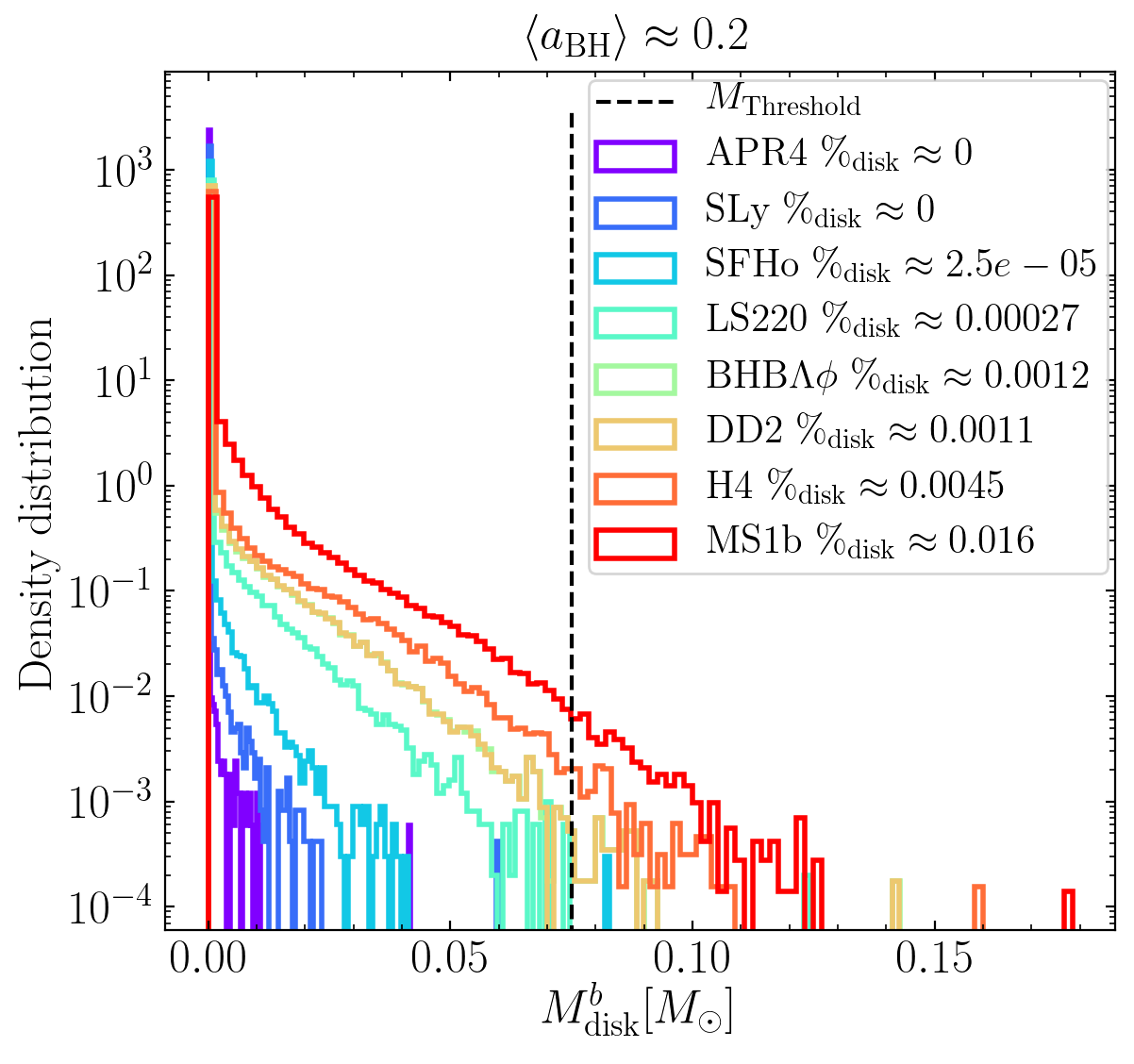}
  \includegraphics[width=0.49\textwidth]{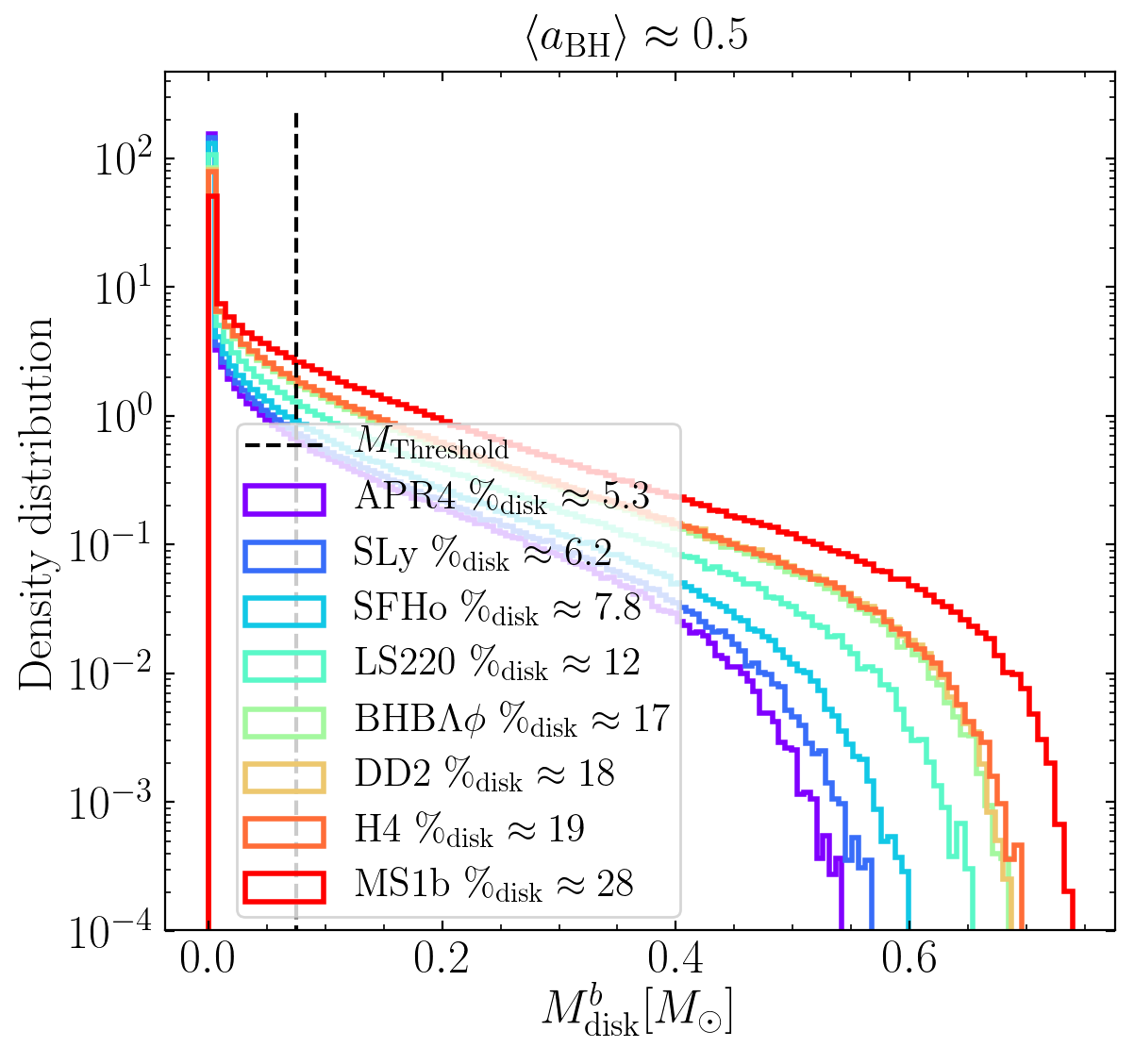}
  \includegraphics[width=0.49\textwidth]{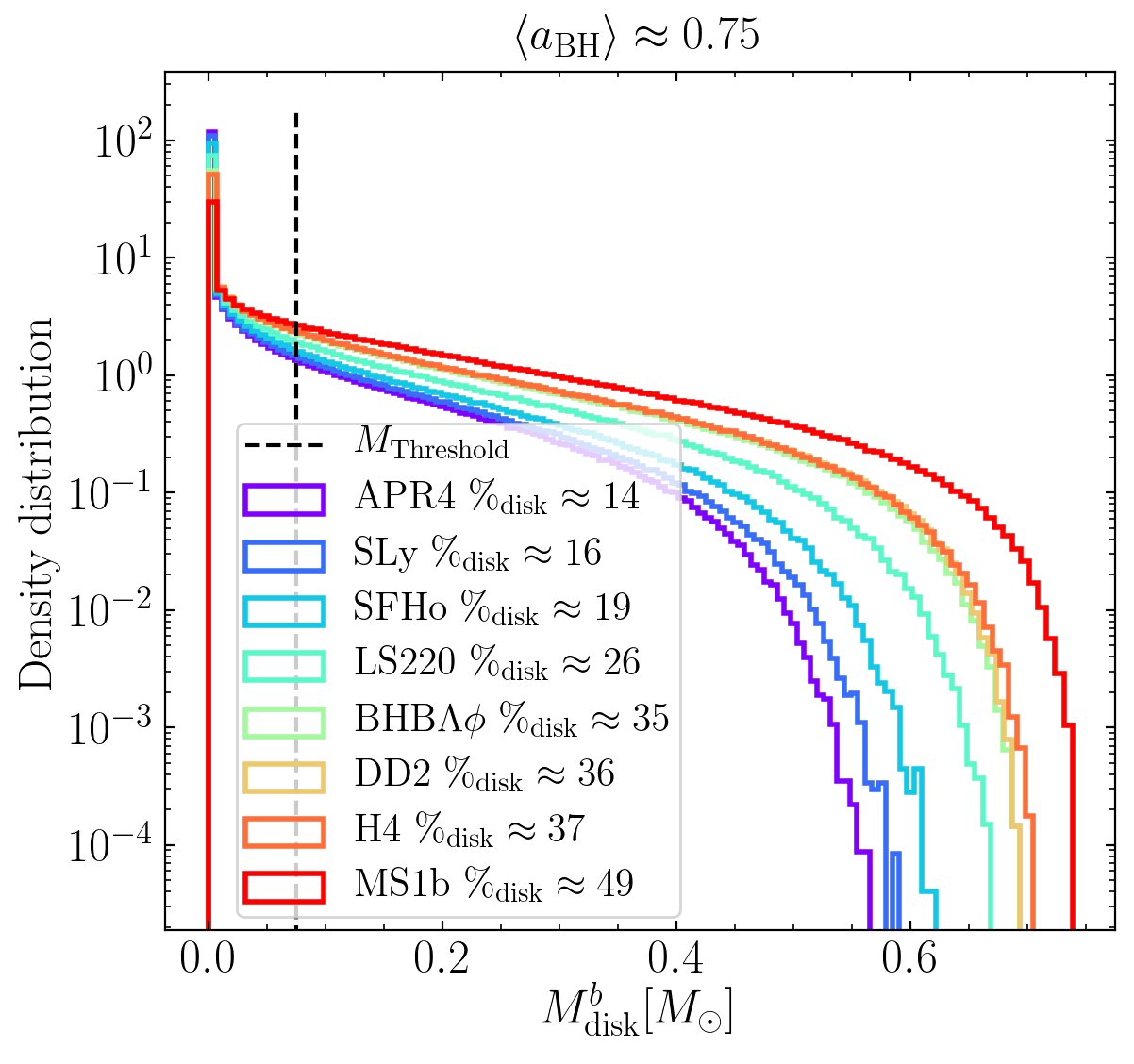}
  \includegraphics[width=0.49\textwidth]{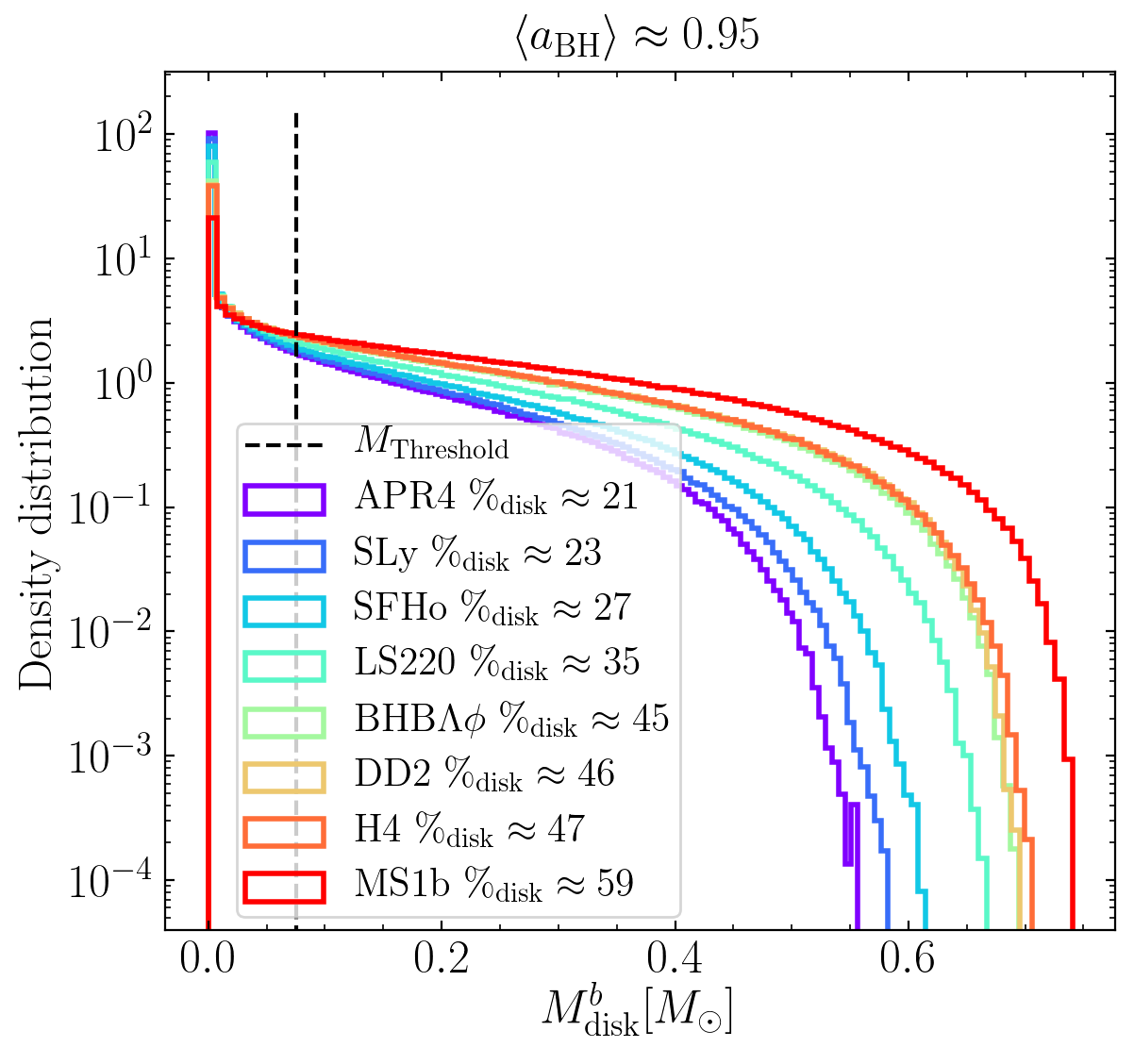}
  \caption{Baryonic mass distributions of the remnant disk for different equations of state 
  		   and under the assumption of aligned BH spins. The mass threshold represents 
  		   the minimum mass of the disk that allows the production of SGRBs with $1$~s duration.
  		   Each panel corresponds to a distribution of the initial BH spin peaked around
  		   the value reported on top. The percentage of binaries with disk mass bigger than the 
  		   threshold is given in the legend for each EOS.
  		   These results have to be considered \textit{upper limits} to the isotropic 
  		   spins case.}
  \label{fig:disk_distr}
\end{figure*}

\autoref{fig:pop_distributions} shows the distribution of the BHNS parameters, 
while \autoref{fig:pop_distributions_rem} shows the distribution of the
remnant masses and spins derived from them. Finally, \autoref{fig:disk_distr} 
gives an overview of the disk analysis considering different distributions
of aligned spins. The disk rest mass is calculated using the model of
\cite{Foucart:2018rjc}.
For the aligned low-spin distribution with $\langle \abhi\rangle=0.2$
no massive disk is formed independently from the EOS. An analogous
result is obtained for our fiducial isotropic spin distributions (not
shown). For the aligned spin distribution with $\langle
\abhi\rangle=0.5$, stiff EOSs give massive disks in $\lesssim20\%$ of
the cases. Note that the MS1b EOS is ruled out by the GW170817
analysis (low-spin priors), while H4 and DD2 are consistent with it. 
Aligned spins populations with $\langle \abhi\rangle\gtrsim0.75$ yield 
more massive disks. Soft EOSs such as APR and SLy admit massive disks in
about $\lesssim20\%$ of the binaries
only if $\langle\abhi\rangle\gtrsim0.95$.


\clearpage 
\begin{center}
\begin{longtable}{cccccccccccccc}
\caption{Numerical relativity simulation data. $M\Omega_0$ refers to the initial orbital
		frequency of the binary system, multiplied by the total mass of the binary system,
		while the other quantities have already been introduced in the first two paragraphs of
		the paper. For some simulations we did not have the final mass and spin of the remnant black hole 
		(dashes in the table), but we could employ them to determine the fitting formula for $L_\text{peak}$.}
\label{tab:data}\\

\hline Name & EOS & $q$ & $\nu$ & 
$M_\text{NS}$ & $a_\text{BH}$ & $M\Omega_0$ & 
$C$ & $M$ & $k_2$ & $\Lambda$ & 
$X_{\bullet}$ & $a_{\bullet}$ & $L_\text{peak}$ \\ \hline 
\endfirsthead

\multicolumn{14}{c}%
{{\bfseries \tablename\ \thetable{} -- continued from previous page}} \\
\hline
Name & EOS & $q$ & $\nu$ & 
$M_\text{NS}$ & $a_\text{BH}$ & $M\Omega_0$ & 
$C$ & $M$ & $k_2$ & $\Lambda$ & 
$X_{\bullet}$ & $a_{\bullet}$ & $L_\text{peak}$ \\ \hline 
\endhead

\hline \multicolumn{14}{|r|}{{Continued on next page}} \\ \hline
\endfoot

\hline
\endlastfoot

2H-Q2M135a75 & 2H & 2 & 0.222 & 1.35 & 0.75 & 0.0250 & 0.1309 & 4.05 & 0.1342 & 2327 & 0.9049 & 0.87 & 2.574e-05\\
1.5H-Q2M135a75 & 1.5H & 2 & 0.222 & 1.35 & 0.75 & 0.0280 & 0.1456 & 4.05 & 0.1190 & 1212 & 0.9094 & 0.89 & 5.002e-05\\
H-Q2M135a75 & H & 2 & 0.222 & 1.35 & 0.75 & 0.0280 & 0.1624 & 4.05 & 0.1031 & 608 & 0.9183 & 0.91 & 9.808e-05\\
HB-Q2M135a75 & HB & 2 & 0.222 & 1.35 & 0.75 & 0.0280 & 0.1718 & 4.05 & 0.0950 & 423 & 0.9242 & 0.91 & 1.400e-04\\
B-Q2M135a75 & B & 2 & 0.222 & 1.35 & 0.75 & 0.0280 & 0.1819 & 4.05 & 0.0863 & 288 & 0.9282 & 0.91 & 1.974e-04\\
2H-Q2M135a5 & 2H & 2 & 0.222 & 1.35 & 0.50 & 0.0250 & 0.1309 & 4.05 & 0.1342 & 2327 & 0.9168 & 0.81 & 2.798e-05\\
1.5H-Q2M135a5 & 1.5H & 2 & 0.222 & 1.35 & 0.50 & 0.0280 & 0.1456 & 4.05 & 0.1190 & 1212 & 0.9262 & 0.82 & 5.481e-05\\
H-Q2M135a5 & H & 2 & 0.222 & 1.35 & 0.50 & 0.0280 & 0.1624 & 4.05 & 0.1031 & 608 & 0.9361 & 0.82 & 1.100e-04\\
HB-Q2M135a5 & HB & 2 & 0.222 & 1.35 & 0.50 & 0.0280 & 0.1718 & 4.05 & 0.0950 & 423 & 0.9421 & 0.83 & 1.594e-04\\
B-Q2M135a5 & B & 2 & 0.222 & 1.35 & 0.50 & 0.0280 & 0.1819 & 4.05 & 0.0863 & 288 & 0.9500 & 0.83 & 2.327e-04\\
2H-Q2M135a-5 & 2H & 2 & 0.222 & 1.35 & -0.50 & 0.0220 & 0.1309 & 4.05 & 0.1342 & 2327 & 0.9536 & 0.48 & 4.196e-05\\
H-Q2M135a-5 & H & 2 & 0.222 & 1.35 & -0.50 & 0.0250 & 0.1624 & 4.05 & 0.1031 & 608 & 0.9770 & 0.51 & 2.023e-04\\
HB-Q2M135a-5 & HB & 2 & 0.222 & 1.35 & -0.50 & 0.0280 & 0.1718 & 4.05 & 0.0950 & 423 & 0.9765 & 0.50 & 2.835e-04\\
B-Q2M135a-5 & B & 2 & 0.222 & 1.35 & -0.50 & 0.0280 & 0.1819 & 4.05 & 0.0863 & 288 & 0.9745 & 0.49 & 3.724e-04\\
2H-Q2M12a75 & 2H & 2 & 0.222 & 1.20 & 0.75 & 0.0250 & 0.1172 & 3.60 & 0.1457 & 4392 & - & - & 1.392e-05\\
H-Q2M12a75 & H & 2 & 0.222 & 1.20 & 0.75 & 0.0280 & 0.1447 & 3.60 & 0.1168 & 1227 & - & - & 4.975e-05\\
HB-Q2M12a75 & HB & 2 & 0.222 & 1.20 & 0.75 & 0.0280 & 0.1527 & 3.60 & 0.1092 & 876 & - & - & 6.712e-05\\
B-Q2M12a75 & B & 2 & 0.222 & 1.20 & 0.75 & 0.0280 & 0.1614 & 3.60 & 0.1012 & 615 & - & - & 9.698e-05\\
2H-Q2M145a75 & 2H & 2 & 0.222 & 1.45 & 0.75 & 0.0250 & 0.1401 & 4.35 & 0.1268 & 1566 & - & - & 3.933e-05\\
H-Q2M145a75 & H & 2 & 0.222 & 1.45 & 0.75 & 0.0280 & 0.1744 & 4.35 & 0.0938 & 387 & - & - & 1.509e-04\\
HB-Q2M145a75 & HB & 2 & 0.222 & 1.45 & 0.75 & 0.0280 & 0.1848 & 4.35 & 0.0854 & 264 & - & - & 2.122e-04\\
B-Q2M145a75 & B & 2 & 0.222 & 1.45 & 0.75 & 0.0280 & 0.1960 & 4.35 & 0.0765 & 176 & - & - & 3.087e-04\\
2H-Q3M135a75 & 2H & 3 & 0.188 & 1.35 & 0.75 & 0.0280 & 0.1309 & 5.40 & 0.1342 & 2327 & 0.9196 & 0.86 & 3.854e-05\\
1.5H-Q3M135a75 & 1.5H & 3 & 0.188 & 1.35 & 0.75 & 0.0300 & 0.1456 & 5.40 & 0.1190 & 1212 & 0.9232 & 0.86 & 7.428e-05\\
H-Q3M135a75 & H & 3 & 0.188 & 1.35 & 0.75 & 0.0300 & 0.1624 & 5.40 & 0.1031 & 608 & 0.9312 & 0.85 & 1.492e-04\\
HB-Q3M135a75 & HB & 3 & 0.188 & 1.35 & 0.75 & 0.0300 & 0.1718 & 5.40 & 0.0950 & 423 & 0.9332 & 0.87 & 2.080e-04\\
B-Q3M135a75 & B & 3 & 0.188 & 1.35 & 0.75 & 0.0300 & 0.1819 & 5.40 & 0.0863 & 288 & 0.9411 & 0.86 & 3.004e-04\\
2H-Q3M135a5 & 2H & 3 & 0.188 & 1.35 & 0.50 & 0.0280 & 0.1309 & 5.40 & 0.1342 & 2327 & 0.9315 & 0.74 & 4.490e-05\\
1.5H-Q3M135a5 & 1.5H & 3 & 0.188 & 1.35 & 0.50 & 0.0300 & 0.1456 & 5.40 & 0.1190 & 1212 & 0.9383 & 0.75 & 9.012e-05\\
H-Q3M135a5 & H & 3 & 0.188 & 1.35 & 0.50 & 0.0300 & 0.1624 & 5.40 & 0.1031 & 608 & 0.9472 & 0.76 & 1.804e-04\\
HB-Q3M135a5 & HB & 3 & 0.188 & 1.35 & 0.50 & 0.0300 & 0.1718 & 5.40 & 0.0950 & 423 & 0.9532 & 0.77 & 2.534e-04\\
B-Q3M135a5 & B & 3 & 0.188 & 1.35 & 0.50 & 0.0300 & 0.1819 & 5.40 & 0.0863 & 288 & 0.9611 & 0.77 & 3.501e-04\\
HB-Q3M135a-5 & HB & 3 & 0.188 & 1.35 & -0.50 & 0.0300 & 0.1718 & 5.40 & 0.0950 & 423 & 0.9785 & 0.32 & 3.007e-04\\
2H-Q4M135a75 & 2H & 4 & 0.160 & 1.35 & 0.75 & 0.0300 & 0.1309 & 6.75 & 0.1342 & 2327 & 0.9303 & 0.84 & 5.202e-05\\
H-Q4M135a75 & H & 4 & 0.160 & 1.35 & 0.75 & 0.0320 & 0.1624 & 6.75 & 0.1031 & 608 & 0.9410 & 0.84 & 1.803e-04\\
HB-Q4M135a75 & HB & 4 & 0.160 & 1.35 & 0.75 & 0.0320 & 0.1718 & 6.75 & 0.0950 & 423 & 0.9459 & 0.85 & 2.527e-04\\
B-Q4M135a75 & B & 4 & 0.160 & 1.35 & 0.75 & 0.0320 & 0.1819 & 6.75 & 0.0863 & 288 & 0.9529 & 0.85 & 3.194e-04\\
2H-Q4M135a5 & 2H & 4 & 0.160 & 1.35 & 0.50 & 0.0350 & 0.1309 & 6.75 & 0.1342 & 2327 & 0.9427 & 0.70 & 6.655e-05\\
H-Q4M135a5 & H & 4 & 0.160 & 1.35 & 0.50 & 0.0350 & 0.1624 & 6.75 & 0.1031 & 608 & 0.9625 & 0.73 & 2.410e-04\\
HB-Q4M135a5 & HB & 4 & 0.160 & 1.35 & 0.50 & 0.0350 & 0.1718 & 6.75 & 0.0950 & 423 & 0.9685 & 0.74 & 3.048e-04\\
B-Q4M135a5 & B & 4 & 0.160 & 1.35 & 0.50 & 0.0350 & 0.1819 & 6.75 & 0.0863 & 288 & 0.9705 & 0.74 & 3.594e-04\\
2H-Q5M135a75 & 2H & 5 & 0.139 & 1.35 & 0.75 & 0.0360 & 0.1309 & 8.10 & 0.1342 & 2327 & 0.9395 & 0.82 & 6.663e-05\\
H-Q5M135a75 & H & 5 & 0.139 & 1.35 & 0.75 & 0.0360 & 0.1624 & 8.10 & 0.1031 & 608 & 0.9534 & 0.84 & 2.081e-04\\
HB-Q5M135a75 & HB & 5 & 0.139 & 1.35 & 0.75 & 0.0360 & 0.1718 & 8.10 & 0.0950 & 423 & 0.9593 & 0.84 & 2.543e-04\\
B-Q5M135a75 & B & 5 & 0.139 & 1.35 & 0.75 & 0.0360 & 0.1819 & 8.10 & 0.0863 & 288 & 0.9652 & 0.85 & 2.945e-04\\
2H-Q2M135 & 2H & 2 & 0.222 & 1.35 & 0.00 & 0.0250 & 0.1309 & 4.05 & 0.1342 & 2327 & 0.9339 & 0.64 & 3.358e-05\\
H-Q2M135 & H & 2 & 0.222 & 1.35 & 0.00 & 0.0280 & 0.1624 & 4.05 & 0.1031 & 608 & 0.9601 & 0.67 & 1.451e-04\\
HB-Q2M135 & HB & 2 & 0.222 & 1.35 & 0.00 & 0.0280 & 0.1718 & 4.05 & 0.0950 & 423 & 0.9691 & 0.67 & 2.215e-04\\
HBs-Q2M135 & HBs & 2 & 0.222 & 1.35 & 0.00 & 0.0280 & 0.1723 & 4.05 & 0.0857 & 376 & 0.9710 & 0.67 & 2.369e-04\\
HBss-Q2M135 & HBss & 2 & 0.222 & 1.35 & 0.00 & 0.0280 & 0.1741 & 4.05 & 0.0724 & 301 & 0.9710 & 0.67 & 2.719e-04\\
B-Q2M135 & B & 2 & 0.222 & 1.35 & 0.00 & 0.0280 & 0.1819 & 4.05 & 0.0863 & 288 & 0.9710 & 0.67 & 3.187e-04\\
Bs-Q2M135 & Bs & 2 & 0.222 & 1.35 & 0.00 & 0.0280 & 0.1856 & 4.05 & 0.0754 & 228 & 0.9710 & 0.66 & 3.694e-04\\
Bss-Q2M135 & Bss & 2 & 0.222 & 1.35 & 0.00 & 0.0280 & 0.1940 & 4.05 & 0.0588 & 142 & 0.9681 & 0.65 & 4.733e-04\\
2H-Q3M135 & 2H & 3 & 0.188 & 1.35 & 0.00 & 0.0280 & 0.1309 & 5.40 & 0.1342 & 2327 & 0.9507 & 0.52 & 6.406e-05\\
H-Q3M135 & H & 3 & 0.188 & 1.35 & 0.00 & 0.0300 & 0.1624 & 5.40 & 0.1031 & 608 & 0.9744 & 0.56 & 2.669e-04\\
HB-Q3M135 & HB & 3 & 0.188 & 1.35 & 0.00 & 0.0300 & 0.1718 & 5.40 & 0.0950 & 423 & 0.9754 & 0.56 & 3.354e-04\\
B-Q3M135 & B & 3 & 0.188 & 1.35 & 0.00 & 0.0300 & 0.1819 & 5.40 & 0.0863 & 288 & 0.9742 & 0.55 & 3.923e-04\\
2H-Q2M12 & 2H & 2 & 0.222 & 1.20 & 0.00 & 0.0220 & 0.1172 & 3.60 & 0.1457 & 4392 & 0.9295 & 0.62 & 1.690e-05\\
H-Q2M12 & H & 2 & 0.222 & 1.20 & 0.00 & 0.0280 & 0.1447 & 3.60 & 0.1168 & 1227 & 0.9492 & 0.66 & 6.877e-05\\
HB-Q2M12 & HB & 2 & 0.222 & 1.20 & 0.00 & 0.0280 & 0.1527 & 3.60 & 0.1092 & 876 & 0.9562 & 0.66 & 9.705e-05\\
B-Q2M12 & B & 2 & 0.222 & 1.20 & 0.00 & 0.0280 & 0.1614 & 3.60 & 0.1012 & 615 & 0.9611 & 0.67 & 1.426e-04\\
APR4-Q3M135a75 & APR4 & 3 & 0.188 & 1.35 & 0.75 & 0.0360 & 0.1800 & 5.40 & 0.0908 & 320 & 0.9389 & 0.87 & 2.527e-04\\
ALF2-Q3M135a75 & ALF2 & 3 & 0.188 & 1.35 & 0.75 & 0.0360 & 0.1610 & 5.40 & 0.1200 & 739 & 0.9296 & 0.86 & 1.278e-04\\
H4-Q3M135a75 & H4 & 3 & 0.188 & 1.35 & 0.75 & 0.0360 & 0.1470 & 5.40 & 0.1150 & 1116 & 0.9241 & 0.88 & 7.523e-05\\
MS1-Q3M135a75 & MS1 & 3 & 0.188 & 1.35 & 0.75 & 0.0360 & 0.1380 & 5.40 & 0.1320 & 1758 & 0.9204 & 0.88 & 5.084e-05\\
APR4-Q3M135a5 & APR4 & 3 & 0.188 & 1.35 & 0.50 & 0.0360 & 0.1800 & 5.40 & 0.0908 & 320 & 0.9574 & 0.77 & 3.068e-04\\
ALF2-Q3M135a5 & ALF2 & 3 & 0.188 & 1.35 & 0.50 & 0.0360 & 0.1610 & 5.40 & 0.1200 & 739 & 0.9444 & 0.76 & 1.487e-04\\
H4-Q3M135a5 & H4 & 3 & 0.188 & 1.35 & 0.50 & 0.0360 & 0.1470 & 5.40 & 0.1150 & 1116 & 0.9389 & 0.76 & 9.326e-05\\
MS1-Q3M135a5 & MS1 & 3 & 0.188 & 1.35 & 0.50 & 0.0360 & 0.1380 & 5.40 & 0.1320 & 1758 & 0.9352 & 0.75 & 6.257e-05\\
APR4-Q3M135 & APR4 & 3 & 0.188 & 1.35 & 0.00 & 0.0360 & 0.1800 & 5.40 & 0.0908 & 320 & 0.9741 & 0.55 & 3.075e-04\\
ALF2-Q3M135 & ALF2 & 3 & 0.188 & 1.35 & 0.00 & 0.0360 & 0.1610 & 5.40 & 0.1200 & 739 & 0.9741 & 0.56 & 2.051e-04\\
H4-Q3M135 & H4 & 3 & 0.188 & 1.35 & 0.00 & 0.0360 & 0.1470 & 5.40 & 0.1150 & 1116 & 0.9630 & 0.55 & 1.419e-04\\
MS1-Q3M135 & MS1 & 3 & 0.188 & 1.35 & 0.00 & 0.0360 & 0.1380 & 5.40 & 0.1320 & 1758 & 0.9556 & 0.53 & 9.185e-05\\
APR4-Q5M135a75 & APR4 & 5 & 0.139 & 1.35 & 0.75 & 0.0400 & 0.1800 & 8.10 & 0.0908 & 320 & 0.9630 & 0.85 & 2.695e-04\\
ALF2-Q5M135a75 & ALF2 & 5 & 0.139 & 1.35 & 0.75 & 0.0400 & 0.1610 & 8.10 & 0.1200 & 739 & 0.9494 & 0.83 & 1.772e-04\\
H4-Q5M135a75 & H4 & 5 & 0.139 & 1.35 & 0.75 & 0.0400 & 0.1470 & 8.10 & 0.1150 & 1116 & 0.9444 & 0.83 & 1.303e-04\\
MS1-Q5M135a75 & MS1 & 5 & 0.139 & 1.35 & 0.75 & 0.0400 & 0.1380 & 8.10 & 0.1320 & 1758 & 0.9407 & 0.83 & 8.638e-05\\
APR4-Q5M135a5 & APR4 & 5 & 0.139 & 1.35 & 0.50 & 0.0400 & 0.1800 & 8.10 & 0.0908 & 320 & 0.9753 & 0.71 & 2.635e-04\\
ALF2-Q5M135a5 & ALF2 & 5 & 0.139 & 1.35 & 0.50 & 0.0400 & 0.1610 & 8.10 & 0.1200 & 739 & 0.9741 & 0.71 & 2.051e-04\\
H4-Q5M135a5 & H4 & 5 & 0.139 & 1.35 & 0.50 & 0.0400 & 0.1470 & 8.10 & 0.1150 & 1116 & 0.9642 & 0.70 & 1.699e-04\\
MS1-Q5M135a5 & MS1 & 5 & 0.139 & 1.35 & 0.50 & 0.0400 & 0.1380 & 8.10 & 0.1320 & 1758 & 0.9556 & 0.68 & 1.181e-04\\
APR4-Q7M135a75 & APR4 & 7 & 0.109 & 1.35 & 0.75 & 0.0440 & 0.1800 & 10.80 & 0.0908 & 320 & 0.9722 & 0.83 & 2.217e-04\\
ALF2-Q7M135a75 & ALF2 & 7 & 0.109 & 1.35 & 0.75 & 0.0440 & 0.1610 & 10.80 & 0.1200 & 739 & 0.9722 & 0.83 & 1.931e-04\\
H4-Q7M135a75 & H4 & 7 & 0.109 & 1.35 & 0.75 & 0.0440 & 0.1470 & 10.80 & 0.1150 & 1116 & 0.9630 & 0.82 & 1.451e-04\\
MS1-Q7M135a75 & MS1 & 7 & 0.109 & 1.35 & 0.75 & 0.0440 & 0.1380 & 10.80 & 0.1320 & 1758 & 0.9537 & 0.81 & 1.063e-04\\
APR4-Q7M135a5 & APR4 & 7 & 0.109 & 1.35 & 0.50 & 0.0440 & 0.1800 & 10.80 & 0.0908 & 320 & 0.9815 & 0.67 & 1.638e-04\\
ALF2-Q7M135a5 & ALF2 & 7 & 0.109 & 1.35 & 0.50 & 0.0440 & 0.1610 & 10.80 & 0.1200 & 739 & 0.9815 & 0.67 & 1.469e-04\\
H4-Q7M135a5 & H4 & 7 & 0.109 & 1.35 & 0.50 & 0.0440 & 0.1470 & 10.80 & 0.1150 & 1116 & 0.9815 & 0.67 & 1.527e-04\\
MS1-Q7M135a5 & MS1 & 7 & 0.109 & 1.35 & 0.50 & 0.0440 & 0.1380 & 10.80 & 0.1320 & 1758 & 0.9815 & 0.67 & 1.393e-04\\
125Hs-Q2M135 & 125Hs & 2 & 0.222 & 1.35 & 0.00 & - & 0.1497 & 4.05 & 0.1051 & 931 & - & - & 8.621e-05\\
Hs-Q2M135 & Hs & 2 & 0.222 & 1.35 & 0.00 & - & 0.1605 & 4.05 & 0.0954 & 597 & - & - & 1.437e-04\\
B-Q4M135 & B & 4 & 0.160 & 1.35 & 0.00 & - & 0.1819 & 6.75 & 0.0861 & 288 & - & - & 2.898e-04\\
Hl-Q2M135 & Hl & 2 & 0.222 & 1.35 & 0.00 & - & 0.1638 & 4.05 & 0.1085 & 613 & - & - & 1.497e-04\\
15H-Q5M135 & 15H & 5 & 0.139 & 1.35 & 0.00 & - & 0.1456 & 8.10 & 0.1189 & 1211 & - & - & 1.852e-04\\
Hss-Q2M135 & Hss & 2 & 0.222 & 1.35 & 0.00 & - & 0.1577 & 4.05 & 0.0850 & 580 & - & - & 1.367e-04\\
B-Q5M135 & B & 5 & 0.139 & 1.35 & 0.00 & - & 0.1819 & 8.10 & 0.0861 & 288 & - & - & 1.932e-04\\
125H-Q2M135 & 125H & 2 & 0.222 & 1.35 & 0.00 & - & 0.1537 & 4.05 & 0.1110 & 862 & - & - & 9.945e-05\\
15H-Q4M135 & 15H & 4 & 0.160 & 1.35 & 0.00 & - & 0.1456 & 6.75 & 0.1189 & 1211 & - & - & 1.901e-04\\
HBl-Q2M135 & HBl & 2 & 0.222 & 1.35 & 0.00 & - & 0.1716 & 4.05 & 0.1013 & 453 & - & - & 1.993e-04\\
125Hl-Q2M135 & 125Hl & 2 & 0.222 & 1.35 & 0.00 & - & 0.1565 & 4.05 & 0.1155 & 820 & - & - & 1.066e-04\\
15Hl-Q2M135 & 15Hl & 2 & 0.222 & 1.35 & 0.00 & - & 0.1497 & 4.05 & 0.1223 & 1084 & - & - & 7.792e-05\\
Bl-Q2M135 & Bl & 2 & 0.222 & 1.35 & 0.00 & - & 0.1798 & 4.05 & 0.0941 & 333 & - & - & 2.688e-04\\
15H-Q2M135 & 15H & 2 & 0.222 & 1.35 & 0.00 & - & 0.1456 & 4.05 & 0.1189 & 1211 & - & - & 6.970e-05\\
15Hs-Q2M135 & 15Hs & 2 & 0.222 & 1.35 & 0.00 & - & 0.1399 & 4.05 & 0.1144 & 1423 & - & - & 5.495e-05\\
15Hss-Q2M135 & 15Hss & 2 & 0.222 & 1.35 & 0.00 & - & 0.1311 & 4.05 & 0.1083 & 1864 & - & - & 3.713e-05\\
H-Q5M135 & H & 5 & 0.139 & 1.35 & 0.00 & - & 0.1625 & 8.10 & 0.1029 & 605 & - & - & 2.052e-04\\
15H-Q3M135 & 15H & 3 & 0.188 & 1.35 & 0.00 & - & 0.1456 & 5.40 & 0.1189 & 1211 & - & - & 1.355e-04\\
125Hss-Q2M135 & 125Hss & 2 & 0.222 & 1.35 & 0.00 & - & 0.1435 & 4.05 & 0.0970 & 1062 & - & - & 6.904e-05\\
2H-Q5M135 & 2H & 5 & 0.139 & 1.35 & 0.00 & - & 0.1310 & 8.10 & 0.1342 & 2319 & - & - & 1.286e-04\\
H-Q4M135 & H & 4 & 0.160 & 1.35 & 0.00 & - & 0.1625 & 6.75 & 0.1029 & 605 & - & - & 2.530e-04\\
2H-Q4M135 & 2H & 4 & 0.160 & 1.35 & 0.00 & - & 0.1310 & 6.75 & 0.1342 & 2319 & - & - & 1.047e-04\\
Hl-Q3M135a5 & Hl & 3 & 0.188 & 1.35 & 0.50 & - & 0.1638 & 5.40 & 0.1085 & 613 & - & - & 1.869e-04\\
125Hs-Q3M135a5 & 125Hs & 3 & 0.188 & 1.35 & 0.50 & - & 0.1497 & 5.40 & 0.1051 & 931 & - & - & 1.126e-04\\
15Hl-Q3M135a5 & 15Hl & 3 & 0.188 & 1.35 & 0.50 & - & 0.1497 & 5.40 & 0.1223 & 1084 & - & - & 1.030e-04\\
HBss-Q3M135a5 & HBss & 3 & 0.188 & 1.35 & 0.50 & - & 0.1741 & 5.40 & 0.0723 & 301 & - & - & 3.136e-04\\
H-Q5M135a5 & H & 5 & 0.139 & 1.35 & 0.50 & - & 0.1625 & 8.10 & 0.1029 & 605 & - & - & 2.285e-04\\
Hss-Q3M135a5 & Hss & 3 & 0.188 & 1.35 & 0.50 & - & 0.1577 & 5.40 & 0.0850 & 580 & - & - & 1.659e-04\\
HBl-Q3M135a5 & HBl & 3 & 0.188 & 1.35 & 0.50 & - & 0.1716 & 5.40 & 0.1013 & 453 & - & - & 2.546e-04\\
125H-Q3M135a5 & 125H & 3 & 0.188 & 1.35 & 0.50 & - & 0.1537 & 5.40 & 0.1110 & 862 & - & - & 1.279e-04\\
125Hss-Q3M135a5 & 125Hss & 3 & 0.188 & 1.35 & 0.50 & - & 0.1435 & 5.40 & 0.0970 & 1062 & - & - & 8.825e-05\\
B-Q5M135a5 & B & 5 & 0.139 & 1.35 & 0.50 & - & 0.1819 & 8.10 & 0.0861 & 288 & - & - & 2.829e-04\\
15H-Q4M135a5 & 15H & 4 & 0.160 & 1.35 & 0.50 & - & 0.1456 & 6.75 & 0.1189 & 1211 & - & - & 1.258e-04\\
Bss-Q3M135a5 & Bss & 3 & 0.188 & 1.35 & 0.50 & - & 0.1941 & 5.40 & 0.0585 & 141 & - & - & 4.242e-04\\
15Hs-Q3M135a5 & 15Hs & 3 & 0.188 & 1.35 & 0.50 & - & 0.1399 & 5.40 & 0.1144 & 1423 & - & - & 7.111e-05\\
Bl-Q3M135a5 & Bl & 3 & 0.188 & 1.35 & 0.50 & - & 0.1798 & 5.40 & 0.0941 & 333 & - & - & 3.284e-04\\
15Hss-Q3M135a5 & 15Hss & 3 & 0.188 & 1.35 & 0.50 & - & 0.1311 & 5.40 & 0.1083 & 1864 & - & - & 4.756e-05\\
15H-Q5M135a5 & 15H & 5 & 0.139 & 1.35 & 0.50 & - & 0.1456 & 8.10 & 0.1189 & 1211 & - & - & 1.665e-04\\
Bs-Q3M135a5 & Bs & 3 & 0.188 & 1.35 & 0.50 & - & 0.1856 & 5.40 & 0.0751 & 227 & - & - & 3.806e-04\\
2H-Q5M135a5 & 2H & 5 & 0.139 & 1.35 & 0.50 & - & 0.1310 & 8.10 & 0.1342 & 2319 & - & - & 8.970e-05\\
HBs-Q3M135a5 & HBs & 3 & 0.188 & 1.35 & 0.50 & - & 0.1723 & 5.40 & 0.0855 & 375 & - & - & 2.597e-04\\
125Hl-Q3M135a5 & 125Hl & 3 & 0.188 & 1.35 & 0.50 & - & 0.1565 & 5.40 & 0.1155 & 820 & - & - & 1.385e-04\\
Hs-Q3M135a5 & Hs & 3 & 0.188 & 1.35 & 0.50 & - & 0.1605 & 5.40 & 0.0954 & 597 & - & - & 1.847e-04\\
B-Q4M135a25 & B & 4 & 0.160 & 1.35 & 0.25 & - & 0.1819 & 6.75 & 0.0861 & 288 & - & - & 3.221e-04\\
B-Q5M135a25 & B & 5 & 0.139 & 1.35 & 0.25 & - & 0.1819 & 8.10 & 0.0861 & 288 & - & - & 2.437e-04\\
H-Q4M135a25 & H & 4 & 0.160 & 1.35 & 0.25 & - & 0.1625 & 6.75 & 0.1029 & 605 & - & - & 2.689e-04\\
15H-Q4M135a25 & 15H & 4 & 0.160 & 1.35 & 0.25 & - & 0.1456 & 6.75 & 0.1189 & 1211 & - & - & 1.668e-04\\
H-Q5M135a25 & H & 5 & 0.139 & 1.35 & 0.25 & - & 0.1625 & 8.10 & 0.1029 & 605 & - & - & 2.275e-04\\
15H-Q3M135a25 & 15H & 3 & 0.188 & 1.35 & 0.25 & - & 0.1456 & 5.40 & 0.1189 & 1211 & - & - & 1.107e-04\\
B-Q3M135a25 & B & 3 & 0.188 & 1.35 & 0.25 & - & 0.1819 & 5.40 & 0.0861 & 288 & - & - & 3.880e-04\\
2H-Q5M135a25 & 2H & 5 & 0.139 & 1.35 & 0.25 & - & 0.1310 & 8.10 & 0.1342 & 2319 & - & - & 1.185e-04\\
15H-Q5M135a25 & 15H & 5 & 0.139 & 1.35 & 0.25 & - & 0.1456 & 8.10 & 0.1189 & 1211 & - & - & 1.893e-04\\
H-Q3M135a25 & H & 3 & 0.188 & 1.35 & 0.25 & - & 0.1625 & 5.40 & 0.1029 & 605 & - & - & 2.228e-04\\
2H-Q3M135a25 & 2H & 3 & 0.188 & 1.35 & 0.25 & - & 0.1310 & 5.40 & 0.1342 & 2319 & - & - & 5.302e-05\\
2H-Q4M135a25 & 2H & 4 & 0.160 & 1.35 & 0.25 & - & 0.1310 & 6.75 & 0.1342 & 2319 & - & - & 7.834e-05\\
Hs-Q5M135a75 & Hs & 5 & 0.139 & 1.35 & 0.75 & - & 0.1605 & 8.10 & 0.0954 & 597 & - & - & 2.164e-04\\
Bss-Q5M135a75 & Bss & 5 & 0.139 & 1.35 & 0.75 & - & 0.1941 & 8.10 & 0.0585 & 141 & - & - & 3.616e-04\\
15H-Q5M135a75 & 15H & 5 & 0.139 & 1.35 & 0.75 & - & 0.1456 & 8.10 & 0.1189 & 1211 & - & - & 1.299e-04\\
HBss-Q5M135a75 & HBss & 5 & 0.139 & 1.35 & 0.75 & - & 0.1741 & 8.10 & 0.0723 & 301 & - & - & 3.009e-04\\
125Hs-Q5M135a75 & 125Hs & 5 & 0.139 & 1.35 & 0.75 & - & 0.1497 & 8.10 & 0.1051 & 931 & - & - & 1.551e-04\\
15Hss-Q5M135a75 & 15Hss & 5 & 0.139 & 1.35 & 0.75 & - & 0.1311 & 8.10 & 0.1083 & 1864 & - & - & 7.046e-05\\
125H-Q5M135a75 & 125H & 5 & 0.139 & 1.35 & 0.75 & - & 0.1537 & 8.10 & 0.1110 & 862 & - & - & 1.706e-04\\
15H-Q4M135a75 & 15H & 4 & 0.160 & 1.35 & 0.75 & - & 0.1456 & 6.75 & 0.1189 & 1211 & - & - & 1.009e-04\\
125Hl-Q5M135a75 & 125Hl & 5 & 0.139 & 1.35 & 0.75 & - & 0.1565 & 8.10 & 0.1155 & 820 & - & - & 1.673e-04\\
15Hs-Q5M135a75 & 15Hs & 5 & 0.139 & 1.35 & 0.75 & - & 0.1399 & 8.10 & 0.1144 & 1423 & - & - & 1.068e-04\\
125Hss-Q5M135a75 & 125Hss & 5 & 0.139 & 1.35 & 0.75 & - & 0.1435 & 8.10 & 0.0970 & 1062 & - & - & 1.283e-04\\
Bs-Q5M135a75 & Bs & 5 & 0.139 & 1.35 & 0.75 & - & 0.1856 & 8.10 & 0.0751 & 227 & - & - & 3.430e-04\\
Hss-Q5M135a75 & Hss & 5 & 0.139 & 1.35 & 0.75 & - & 0.1577 & 8.10 & 0.0850 & 580 & - & - & 2.099e-04\\
Bl-Q5M135a75 & Bl & 5 & 0.139 & 1.35 & 0.75 & - & 0.1798 & 8.10 & 0.0941 & 333 & - & - & 3.077e-04\\
HBs-Q5M135a75 & HBs & 5 & 0.139 & 1.35 & 0.75 & - & 0.1723 & 8.10 & 0.0855 & 375 & - & - & 2.775e-04\\
Hl-Q5M135a75 & Hl & 5 & 0.139 & 1.35 & 0.75 & - & 0.1638 & 8.10 & 0.1085 & 613 & - & - & 2.139e-04\\
HBl-Q5M135a75 & HBl & 5 & 0.139 & 1.35 & 0.75 & - & 0.1716 & 8.10 & 0.1013 & 453 & - & - & 2.576e-04\\
15Hl-Q5M135a75 & 15Hl & 5 & 0.139 & 1.35 & 0.75 & - & 0.1497 & 8.10 & 0.1223 & 1084 & - & - & 1.356e-04\\

\end{longtable}
\end{center}
\end{document}